\input epsf

\magnification\magstep1




\overfullrule=0pt  
\hbadness=10000      
\vbadness=10000

\font\eightit=cmti8  
\font\eightrm=cmr8 \font\eighti=cmmi8                 
\font\eightsy=cmsy8 
\font\sixrm=cmr6


\def\eightpoint{\normalbaselineskip=10pt 
\def\rm{\eightrm\fam0} \let\it\eightit
\textfont0=\eightrm \scriptfont0=\sixrm 
\textfont1=\eighti \scriptfont1=\seveni
\textfont2=\eightsy \scriptfont2=\sevensy 
\normalbaselines \eightrm
\parindent=1em}



\def\eq#1{{\noexpand\rm(#1)}}          
\newcount\eqcounter                    
\eqcounter=0                           
\def\numeq{\global\advance\eqcounter by 1\eq{\the\eqcounter}}           
\def\relativeq#1{{\advance\eqcounter by #1\eq{\the\eqcounter}}}



\def\namelasteq#1{\global\edef#1{{\eq{\the\eqcounter}}}}  


\def\A{{\rm A}}

 
\def\B{{\rm B}}                      
\def\cite#1{{\rm[#1]}}                 

\def\eps{\varepsilon}                  


\def\ghat{\hat g}                      
\def\gbar{\bar g}                      
\def\gamhat{\hat\gamma}                
\def\gambar{\bar\gamma}                
\def\gam5{\gamma_5}                    
\def\io{i_1}
\def\id{i_2}
\def\ih{i_3}
\def\jo{j_1}
\def\jd{j_2}
\def\jh{j_3}
\def\intd{\int\! {\rm d}}              
\def\nocorr{\kern0pt}                  
\def\pslash{{p\mkern-8mu/}{\!}}        
\def\poneslash{{p\mkern-8mu/}_1{\!}} 
\def\ptwoslash{{p\mkern-8mu/}_2{\!}}

\def\prslash{{\partial\mkern-9mu/}}    
\def\qslash{{q\mkern-8mu/}{\!}}
\def\T{{\rm T}}

\def\Taone{\T^{a_1}}
\def\Tatwo{\T^{a_2}}
\def\Tathree{\T^{a_3}}
\def\Tr{{\rm Tr}}                      


\def\teta{{1\over 2}\theta}
\def\iteta{{i\over 2}\theta}
\def\eps{\rm eps}
\def\cite#1{{\rm[#1]}}                 



\newif\ifstartsec                   

\outer\def\section#1{\vskip 0pt plus .15\vsize \penalty -250
\vskip 0pt plus -.15\vsize \bigskip \startsectrue
\message{#1}\centerline{\bf#1}\nobreak\noindent}

\def\subsection#1{\ifstartsec\medskip\else\bigskip\fi \startsecfalse
\noindent{\it#1}\penalty100\medskip}

\def\refno#1. #2\par{\smallskip\item{\rm\lbrack#1\rbrack}#2\par}

\def\elementos{1}
\def\AGM{2}
\def\Terashima{3}
\def\anomalPP{4}
\def\Bonora{5}
\def\Adler{6}
\def\ColemanGrossman{7}
\def\MRS{8}
\def\Intanom{9}
\def\BreitMaison{10}
\def\Pernici{11}
\def\gammafive{12}
\def\AGO{13}
\def\RT{14}
\def\Bonoraetal{15}
\def\Arsa{16}
\def\Gravity{17}


\rightline{FT/UCM--50--2000}

\vskip 1cm

\centerline{\bf  The UV and IR Origin  of  Nonabelian Chiral Gauge}
\centerline{\bf  Anomalies on  Noncommutative Minkowski Space-time}

\bigskip 

\centerline{\rm C. P.
Mart{\'\i}n${}^{\dag}$}
\medskip
\centerline{\eightit Departamento de F{\'\i}sica Te\'orica I,				
		Universidad Complutense, 28040 Madrid, Spain}
\vfootnote\dag{email: {\tt carmelo@elbereth.fis.ucm.es}}

\bigskip\bigskip

\begingroup\narrower\narrower
\eightpoint
We discuss both the  UV and IR  origin of the one-loop triangle gauge  
anomalies  for noncommutative   nonabelian chiral gauge theories with 
fundamental, adjoint and bi-fundamental fermions for $U(N)$ groups.   
We find that gauge anomalies only come from planar 
triangle diagrams, the non-planar triangle contributions  giving rise
to no breaking of the Ward identies.  Generally speaking, theories with 
fundamental  and bi-fundamental chiral  matter are anomalous.  
Theories  with only adjoint chiral fermions are  
anomaly free. 
\par
\endgroup 
\vskip 2cm

\section{1. Introduction}

Let space-time be noncommutative~\cite{\elementos} Minkowski and 
let $\psi$ denote a fermion chirally coupled to a $U(N)$ gauge 
field $\A_{\mu}$. Let $\A_{\mu}$ be an $N\times N$ matrix which transforms 
under an infinitesimal gauge transformation as follows 
$$
\big(\delta_{\omega} \A_{\mu}\big)^i_{\;j} = 
\partial_{\mu}\omega^i_{\;j}-i\,\A^i_{\mu\,k}\star\omega^k_{\;j}+
i\,\omega^i_{\;k}\star\A^k_{\mu\,j},
\eqno\numeq\namelasteq\Afieldtrans
$$
where 
$\omega^i_{\;j}=\omega^{*\,j}_{\quad i}$, $i,j=1,\cdots, N$, 
are the infinitesimal gauge  transformation parameters and the symbol $\star$ 
stands for the Moyal product of functions on Minkowski space-time.
The Moyal product is defined thus
$$
(f\star g)(x)= 
e^{{i\over 2}\theta^{\mu\nu}\partial_{\mu}^{u}\partial_{\nu}^{w}}
f(u)g(w)\mid_{u=x,w=x},
$$ 
where $\theta^{\mu\nu}$ is an antisymmetric real matrix  either of  
magnetic type or light-like type~\cite{\AGM}. 

Following ref.~\cite{\Terashima}, we introduce three basic right-handed 
chiral gauge transformation laws for the fermion field
$$
(\delta_{\omega}\psi)^i=i\,\omega^i_{\;j}\star {\rm P}_{+}\,\psi^j\qquad{\rm and}
\qquad
(\delta_{\omega} \bar{\psi})_{k}=
-i\bar{\psi}_k\star\omega^{k}_{\;i}{\rm P}_{-}, 
\eqno\numeq\namelasteq\fundamental
$$
$$
(\delta_{\omega}\psi)_j= -i\,{\rm P}_{+}\,\psi_i\star
\omega^i_{\;j}\qquad{\rm and}
\qquad
(\delta_{\omega} \bar{\psi})^k=
i \omega^{k}_{\;i}\star\bar{\psi}^i\,{\rm P}_{-}
\eqno\numeq\namelasteq\antifundamental
$$
and
$$
(\delta_{\omega}\psi)^i_{\,j}=
i\,\bigg(\omega^i_{\;j}\star {\rm P}_{+}\,\psi^j_{\; i}
-{\rm P}_{+}\,\psi^i_{\;j}\star\omega^j_{\;i}\bigg)\qquad{\rm and}
\qquad
(\delta_{\omega} \bar{\psi})^k_{\,i}=
-i\bigg(\bar{\psi}^k_{\; j}\star\omega^{j}_{\;i}\,{\rm P}_{-} 
-\omega^{k}_{\;j}\star\bar{\psi}^j_{\;i}\,{\rm P}_{-}\bigg).
\eqno\numeq\namelasteq\adjoint
$$
As usual, ${\rm P}_{+}={1\over 2}(1+\gamma_5)$. The fermions transforming under 
gauge transformations as in eqs.~\fundamental ,~\antifundamental\ and
\adjoint\    will be called (right-handed) fundamental, 
(right-handed) anti-fundamental and (right-handed) 
adjoint fermions, respectively. 

The $U(N)$ chiral gauge theories with the  fermion $\psi$ transforming as in 
eqs.~\fundamental,~\antifundamental\ and \adjoint\ are governed, 
respectively, by  the following classical actions
$$
{\rm S}\;=\;\intd^{4}x\; \bar{\psi}_{i}\star(i\prslash\psi^{i} +
\A^{i}_{\mu\,j}\star\gamma^{\mu} {\rm P}_{+}\psi^{j}),
\eqno\numeq\namelasteq\fundamentalaction
$$ 
$$
{\rm S}\;=\;\intd^{4}x\; \bar{\psi}^{i}\star(i\prslash\psi_{i} - 
\gamma_{\mu}{\rm P}_{+}\psi_{j}\star \A^{j}_{\mu\,i}),
\eqno\numeq\namelasteq\antifundamentalaction
$$ 
and
$$
{\rm S}\;=\;\intd^{4}x\; \bar{\psi}^{k}_{\;i}\star(i\prslash\psi^{i}_{\;k}
+ \A^{i}_{\mu\,j}\star \gamma^{\mu}{\rm P}_{+}\psi^{j}_{\;k}
- \gamma^{\mu}{\rm P}_{+}\psi^{i}_{\;j}\star \A^{j}_{\mu\,k}).
\eqno\numeq\namelasteq\adjointaction
$$ 
Each action is invariant under the corresponding chiral gauge transformations;
these transformations are displayed in 
eqs.~\Afieldtrans,~\fundamental,~\antifundamental\ and \adjoint.

The effective action, $\Gamma[\A]$, which arises upon integrating out the
fermionic degrees of freedom is formally given by 
$$
 e^{i\Gamma[\A]}\;=\;\intd \psi d\bar{\psi}\;
e^{iS[A,\psi, \bar{\psi}]},
\eqno\numeq\namelasteq\effectiveaction
$$
with $S[A,\psi, \bar{\psi}]$ given by any of the classical actions in 
eqs.~\fundamentalaction,~\antifundamentalaction\ and \adjointaction.
The path integral above is formally invariant under the corresponding chiral 
gauge transformations  -see eqs.~\Afieldtrans,~\fundamental,~
\antifundamental\ and \adjoint,
which leads, formally, to the gauge invariance of $\Gamma[\A]$. And yet,
it has been shown in ref.~\cite{\anomalPP} that once the path integral is 
properly defined {\it \'a la } Berezin the effective action is no longer
gauge invariant for fermions transforming as in eqs.~\fundamental ,
\antifundamental , but rather the following anomaly equation  holds
$$
\delta_{\theta}\Gamma[A]\,=\,
\pm\,{1\over 24\pi^2}\,{\rm Tr}\int d^4 x\,
\varepsilon^{\mu_1\mu_2\mu_3\mu_4}\,\theta\,\partial_{\mu_1}\bigl[
\A_{\mu_2}\star\partial_{\mu_3}\A_{\mu_4}-
{i\over 2}\A_{\mu_2}\star\A_{\mu_3}\star\A_{\mu_4}\bigr].
\eqno\numeq\namelasteq\fanomaly
$$
Where the overall $+$ and $-$ signs are for right-handed   
fundamental and  right-handed anti-fundamental fermions, respectively. 
This equation can also be obtained by using standard diagrammatic techniques.
One begins by working out the anomaly equation for 
the three-point contribution -the famous triangle diagrams- to 
$\Gamma[\A]$, the latter has been defined in eq.~{\effectiveaction}, and, then,
one uses the Wess-Zumino consistency condition~\cite{\Bonora} to obtain the
complete equation. Agreement with eq.~\fanomaly\ demands that this triangle
anomaly reads
$$
\eqalignno{
&p_3^{\mu_3}\,\Gamma_{\mu_1\mu_2\mu_3}^{a_1 a_2 a_3}(p_1,p_2)^{\eps}=
\mp{1 \over 24\pi^2}\,\varepsilon_{\mu_1\mu_2\alpha\beta}\, 
p_1^{\alpha}p_2^{\beta}\cr
&\bigg({\rm Tr}\,\{\Taone,\Tatwo\}\,\Tathree\cos \teta(p_1,p_2)
-i{\rm Tr}\,[\Taone,\Tatwo]\Tathree\sin\teta(p_1,p_2)\bigg),
&\numeq\cr
}
\namelasteq{\triangleanomaly}
$$
where $\Gamma_{\mu_1\mu_2\mu_3}^{a_1 a_2 a_3}(p_1,p_2)$ gives the 
Fourier transform, 
$$
\Gamma_{\mu_1\mu_2\mu_3}^{a_1 a_2 a_3}(p_1,p_2,p_3)=
(2\pi)^4\delta(p_1+p_2+p_3)\,\Gamma_{\mu_1\mu_2\mu_3}^{a_1 a_2 a_3}(p_1,p_2),
$$
of the three-point function
$$
{\delta^3 i\Gamma[\A] \over 
\delta \A^{a_1}_{\mu_1}(x_1) 
\delta \A^{a_2}_{\mu_2}(x_2) \delta \A^{a_3}_{\mu_3}(x_3)}
\bigg|_{A=0}\,=\,\int\,\prod_{i=1}^{3} {d^4 p_i\over (2\pi)^4}\; e^{ip_i x}\;
\Gamma_{\mu_1\mu_2\mu_3}^{a_1 a_2 a_3}(p_1,p_2,p_3), 
\eqno\numeq\namelasteq\threepointfunction
$$ 
and the superscript $\eps$ stands for the contribution to
this Green function which carries the Levi-Civita pseudotensor. The indices 
$a_1, a_2$ and $a_3$ run over the generators of the gauge group. The symbol 
$\theta(p_1,p_2)$ is a shorthand for 
$p_{1\,{\mu}}\,\theta^{\mu\nu}\,p_{2\,{\nu}}$. Eq.~\triangleanomaly\
leads clearly  to the conclusion that the triangle contribution on 
noncommutative Minkowski for (anti-) fundamental chiral fermions 
is anomaly free 
if, and only if, 
$$
 {\rm Tr}\,\Taone\Tatwo\Tathree=0;
$$
its ordinary counterpart being
${\rm Tr}\,\{\Taone,\Tatwo\}\Tathree=0$.

We can also have chiral gauge theories with bi-fundamental chiral fermions 
$\psi^{i}_{R\,j}={\rm P}_{+}\psi^{i}_{\,j}$, $i=1,\cdots, N$ and $j=1,\cdots, M$~
\cite{\Terashima}. 
Now the fermion couples to a $U(N)$ gauge field, say,  $\A_{\mu}$ 
and a $U(M)$ gauge field, say, $\B_{\mu}$, the former being an $N\times N$ 
matrix  and the latter an $M\times M$ matrix. The classical action for this
theory reads
$$
{\rm S}\;=\;\intd^{4}x\; \bar{\psi}^{k}_{\;i}\star(i\prslash\psi^{i}_{\;k}
+ \A^i_{\mu\,j}\star \gamma^{\mu}{\rm P}_{+}\psi^{j}_{\;k}
- \gamma^{\mu}{\rm P}_{+}\psi^{i}_{\;j}\star {\B}^j_{\mu\,k}),
\eqno\numeq\namelasteq\biaction
$$ 
This action is invariant under the following infinitesimal gauge  
transformations
$$
\eqalign{&
(\delta_{(\omega,\chi)} \psi)^i_{\,j}=
i\bigg(\omega^{i}_{\;j}\star {\rm P}_{+}\psi^j_{\; i}
-{\rm P}_{+}\psi^i_{\;j}\star\chi^{j}_{\;i}\bigg),\cr
&(\delta_{(\omega,\chi)} \bar{\psi})^k_{\,i}=
-i\bigg(\bar{\psi}^k_{\; j}\star\omega^{j}_{\;i}{\rm P}_{-} 
-\chi^{k}_{\;j}\star\bar{\psi}^j_{\;i}{\rm P}_{-}\bigg),\cr
&
\big(\delta_{\omega} \A_{\mu}\big)^i_{\;j} = 
\partial_{\mu}\omega^i_{\;j}-i\,\A^i_{\mu\,k}\star\omega^k_{\;j}+
i\,\omega^i_{\;k}\star\A^k_{\mu\,j},\cr
&
\big(\delta_{\chi} {\B}_{\mu}\big)^i_{\;j} = 
\partial_{\mu}\chi^i_{\;j}-i\,{\B}^i_{\mu\,k}\star\chi^k_{\;j}+
i\,\chi^i_{\;k}\star{\B}^k_{\mu\,j},\cr
} 
$$
where 
$\omega^i_{\;j}=\omega^{*\,j}_{\quad i}$, $i,j=1,\cdots,N$, and 
$\chi^i_{\;j}=\chi^{*\,j}_{\quad i}$, $i,j=1,\cdots,M$, are the infinitesimal 
gauge transformation parameters.

The effective action, $\Gamma[\A, \B]$, that one obtains by integrating over 
the fermionic degrees of freedom  formally reads thus
$$
 e^{i\Gamma[\A,\B]}\;=\;\intd \psi d\bar{\psi}\;
e^{iS[\A,\B,\psi, \bar{\psi}]},
\eqno\numeq\namelasteq\bieffectiveaction
$$
with $S[\A,\B,\psi, \bar{\psi}]$ given in eq.~\biaction. We shall see that
in general there are triangle gauge anomalies jeopardizing the formal
gauge invariance of  $\Gamma[\A, \B]$.

It is well known that the chiral gauge anomaly on ordinary 
Minkowski space-time  can be understood either as a 
short-distance phenomenon (UV effect)~\cite{\Adler} 
or as an IR effect (large-distance phenomenon)~\cite{\ColemanGrossman}. The
purpose of this paper is to show that  nonabelian chiral anomalies on 
noncommutative Minkowski space-time can also be explained either 
as an UV effect  or an IR phenomenon. Recall that if the chiral fermions 
of the theory are either adjoint or bi-fundamental, there are non-planar 
contributions to the three-point function of the effective action 
($\Gamma_{\rm adj}[\A]$ and 
$\Gamma[\A,\B]$ in eqs.~\effectiveaction\ and \bieffectiveaction)  and one
wonders whether these non-planar contributions  may give rise to 
some gauge anomaly due to its noncommutative  IR structure; 
this structure being a consequence of their being regularized in the UV 
by the appropriate Moyal exponentials~\cite{\MRS}. 
We shall show in this paper that, at least for the theories we have studied,
there are no  anomalous contributions coming from the
nonplanar triangle diagrams: gauge anomalies -if they exist- are due
to  planar triangle diagrams. 
We have assumed that, as in the ordinary case, true anomalies always 
involve the Levy-Civita pseudotensor. Standard arguments~\cite{\Intanom} 
can put forward to support this assumption. 

The layout of this paper is as follows.   
Section 2 is devoted the
analysis of the  of the anomaly equation -eq.~\triangleanomaly- as an UV 
effect. In this section we will also show that the chiral gauge theory whose 
classical action given in eq.~\adjointaction\ is anomaly free.  We close the
section by computing the triangle gauge anomalies for a chiral theory with a
bi-fundamental right-handed fermion and conclude that they only come from the
planar contribution to its effective action; the nonplanar part being thus
anomaly free. In section 3 we shall exhibit the IR origin of the nonabelian 
chiral anomalies we have worked out in section 2. 
We include an Appendix with the relevant Feynman integrals.

\section{2. The UV origin of nonabelian chiral gauge anomalies}

Let us begin with the chiral theory whose action is given by 
eq.~\fundamentalaction. The UV character of eq.~\triangleanomaly\ 
is made apparent by computing its
l.h.s with the help of a regularization method. We shall use dimensional
regularization as defined by 
Breitenlohner and Maison~\cite{\BreitMaison} (see ref.~\cite{\Pernici} for an
alternative), i.e., with the definition of 
$\gamma_5$ given by 't Hooft and Veltman, and take the following 
classical action in the ``d-dimensional'' 
space of dimensional regularization (see ref.~\cite{\gammafive} and references
therein):
$$
{\rm S}\;=\;\intd^{d}x\; \bar{\psi}_{i}\star(i\prslash\psi^{i} + 
A^a_{\mu}T^{a\,i}_{\quad j}\gambar^{\mu}{\rm P}_{+}\star\psi^{j}).
$$  
Here, $T^{a\,i}_{\quad j}=T^{*\,a\,j}_{\;\;\,\quad i}$. The object denoted by 
the symbol $\gambar^{\mu}$ and the other objects in the  algebra of ``d-dimensional'' covariants are defined as in section
2. of ref.~\cite{\gammafive}. The ``d-dimensional'' counterpart of 
$\theta^{\mu\nu}$ is defined as an object which satisfies
$$
\theta^{\mu\nu}=-\theta^{\nu\mu},\quad
\hat{g}_{\mu\rho}\theta^{\rho\nu}=0,\quad 
p_{\mu}\,\theta^{\mu\rho}\eta_{\rho\sigma}\theta^{\sigma\nu}\,p_{\nu}
\geq 0,\; \forall p_{\mu}.
$$ 
The Feynman rules needed to reproduce our computations are given in figure 1.

\topinsert

{\settabs 4\columns \def\graphwidth{1.1in} 
\eightpoint

\+&
 \hfil$\vcenter{\epsfxsize=\graphwidth\epsffile{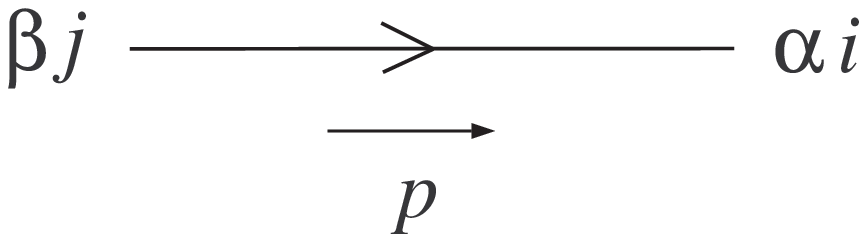}}$\hfil
 & $\vcenter{
     \hbox{${i\; \pslash_{\alpha\beta}\;\delta^{i}{\,j}\over p^2 + i0^{+}}      
           $}
     \vskip 9pt
     }$\cr
\bigskip
\medskip

\+&
 \hfil$\vcenter{\epsfxsize=\graphwidth\epsffile{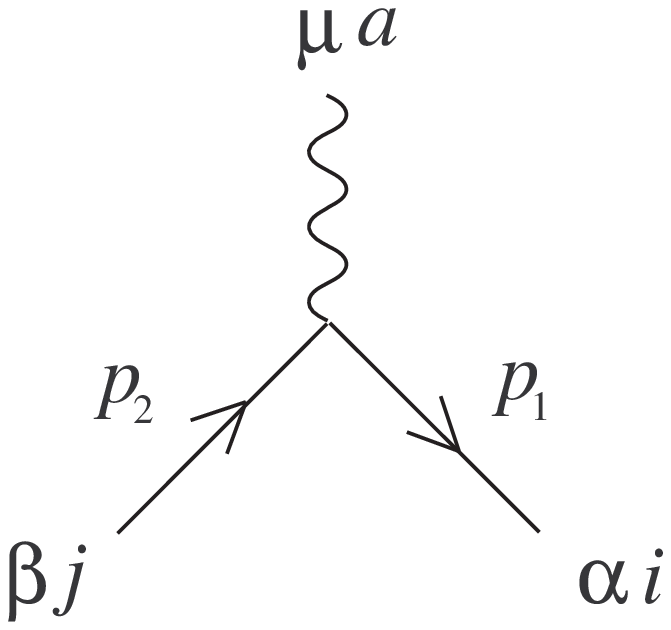}}$\hfil
 & $\vcenter{
      \hbox{$i\; e^{-\iteta(p_1,p_2)}\;T^{a\,i}{\,j}\,
             (\bar\gamma^\mu\,{\rm P}_{+})_{\alpha\beta} 
            $}
  }$\cr
}
\vskip 12pt
\centerline{ {\bf Figure 1.}
      {\eightpoint 
      }
}


\vskip 0.1cm
\endinsert

Let us define the dimensionally regularized counterpart of the l.h.s of
eq.~\triangleanomaly:
$$
\triangle^{a_1 a_2 a_3}_{\mu_1\mu_2}(p_1,p_2;d)=
p_3^{\mu_3}\,\Gamma_{\mu_1\mu_2\mu_3}^{a_1 a_2 a_3}(p_1,p_2;d)^{\eps}.
$$

At the one-loop level $\triangle^{a_1 a_2 a_3}_{\mu_1\mu_2}(p_1,p_2;d)$
is given by the sum of the contributions coming from the two triangle diagrams
in figure 2. This sum reads
$$
\eqalignno{\triangle^{a_1 a_2 a_3}_{\mu_1\mu_2}(p_1,p_2;d)=
&e^{-\iteta(p_1,p_2)}\,{\rm Tr}\,\Taone\Tatwo\Tathree
\,\triangle^{(1)}_{\mu_1\mu_2}(p_1,p_2;d)+\cr
&e^{\iteta(p_1,p_2)}\,{\rm Tr}\,\Tatwo\Taone\Tathree
\,\triangle^{(2)}_{\mu_1\mu_2}(p_1,p_2;d), 
&\numeq\cr
}
\namelasteq{\trianglegraphs}
$$
with
$$
\eqalignno{&\triangle^{(1)}_{\mu_1\mu_2}(p_1,p_2;d)\,=\,
-\,\int{d^d q\over(2\pi)^d}
{{\rm tr}^{\,\eps}\big\{ (\qslash+\poneslash)\,\gambar_{\mu_1}{\rm P}_{+}\,
\qslash\,\gambar_{\mu_2}{\rm P}_{+}\,(\qslash-\ptwoslash)(\bar{\poneslash}+
\bar{\ptwoslash}){\rm P}_{+}\big\}
\over (q^2+i0^+) ((q+p_1)^2+i0^+) ((q-p_2)^2+i0^+)},\cr
&{\rm and}\cr
&\triangle^{(2)}_{\mu_1\mu_2}(p_1,p_2;d)\,=\,
-\,\int{d^d q\over(2\pi)^d}
{{\rm tr}^{\,\eps}\big\{ (\qslash+\ptwoslash)\,\gambar_{\mu_2}{\rm P}_{+}\,
\qslash\,\gambar_{\mu_1}{\rm P}_{+}\,(\qslash-\poneslash)(\bar{\poneslash}+
\bar{\ptwoslash}){\rm P}_{+}\big\}
\over (q^2+i0^+) ((q+p_2)^2+i0^+) ((q-p_1)^2+i0^+)}.
&\numeq\cr
} 
\namelasteq{\contractedintegrals}
$$
In the previous equation ${\rm tr}^{\,\eps}$ shows that only contributions
involving the Levi-Civita symbol $\varepsilon_{\mu_1\mu_2\mu_3\mu_4}$ are kept
upon computing the trace over the gammas.

\topinsert

{\settabs 4\columns \def\graphwidth{1.2in}   
\eightpoint
\+&\hfill\epsfxsize=\graphwidth\epsffile{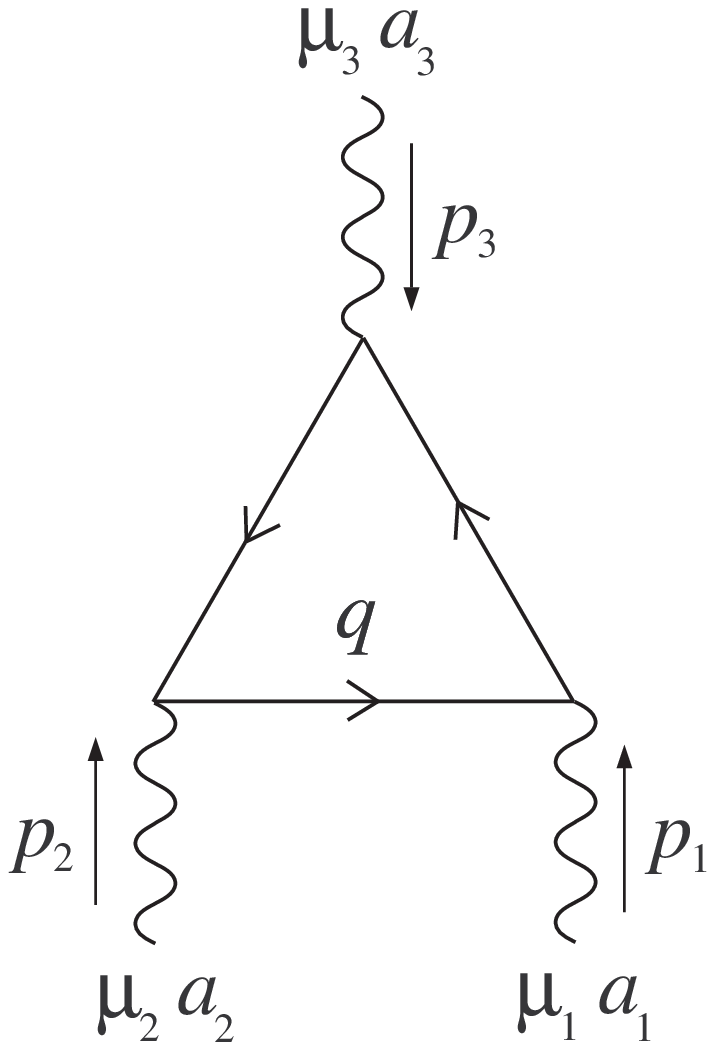}\hfill
	&\hfill\epsfxsize=\graphwidth\epsffile{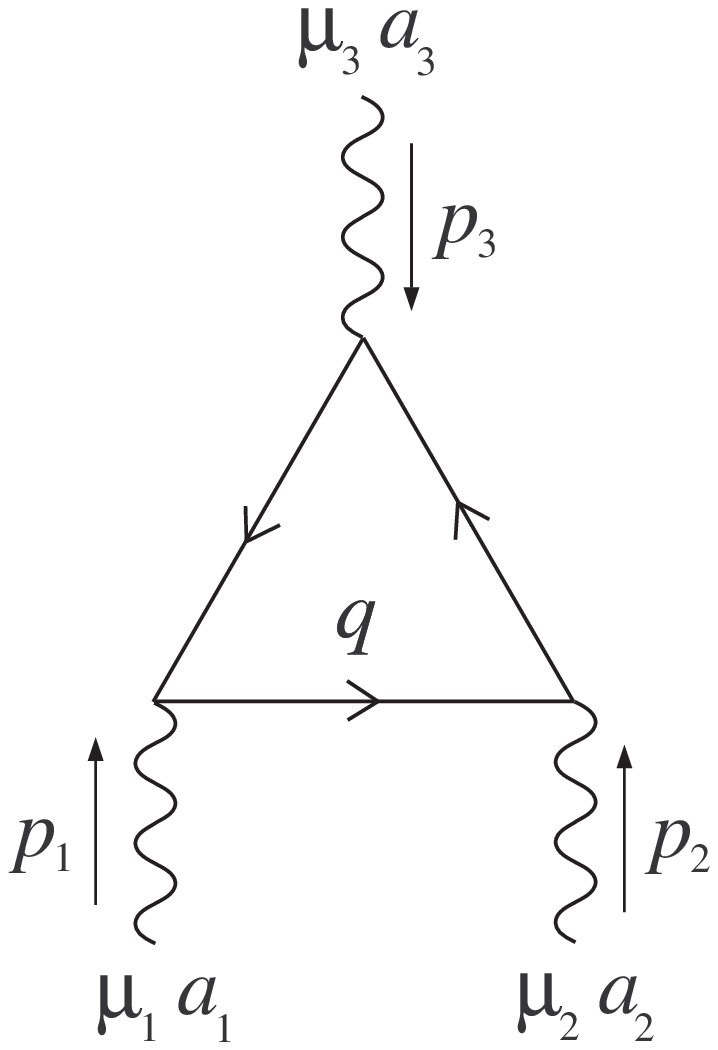}\hfill
	& &\cr
\+&\hfill$(i)$\hfill
	&\hfill $(ii)$\hfill
	& &\cr
}

\vskip 12pt

\centerline{ {\bf Figure 2.}
    {\eightpoint 
    }
}


%
%
\vskip 0.4cm
\endinsert

 Now, the Feynman diagrams in figure 2 are planar; hence,  
it can be readily seen~\cite{\AGO} that their noncomutative 
character is completely embodied (see eq.~\trianglegraphs)  
in the overall phase factors $e^{-\iteta(p_1,p_2)}$ and 
$e^{\iteta(p_1,p_2)}$. Then, it does not come as 
a surprise that equation~\triangleanomaly\ holds, for
the Feynman integrals in eq.~\contractedintegrals\ are the standard
integrals whose UV behaviour give rise to the nonabelian chiral anomaly on
commutative Minkowski space.

Taking into account that ${\rm P}_+\gamhat_{\mu}\gambar_{\nu}{\rm P}_+ =0$ and 
performing some standard manipulations one shows that
$$
\eqalignno{&\triangle^{(1)}_{\mu_1\mu_2}(p_1,p_2;d)\,=\,\cr
&-{1\over 2}\,\int{d^d q\over(2\pi)^d}
{{\rm tr}\big\{ \gambar_{\alpha} \gambar_{\mu_1}\gambar_{\beta}
\gambar_{\mu_2}\gamma_5\big\}\,
\bar{p}_1^{\alpha}(\bar{q}+\bar{p}_2)^{\beta}\,\hat{q}^2\over
(q^2+i0^+) ((q+p_2)^2+i0^+) ((q+p_1+p_2)^2+i0^+)}\cr
&-{1\over 2}\,\int{d^d q\over(2\pi)^d}
{{\rm tr}\big\{ \gambar_{\mu_1} \gambar_{\alpha}\gambar_{\mu_2}
\gambar_{\beta}\gamma_5\big\}\,
(\bar{q}+\bar{p}_1)^{\alpha}\bar{p}_2^{\beta}\,\hat{q}^2\over
(q^2+i0^+) ((q+p_1)^2+i0^+) ((q+p_1+p_2)^2+i0^+)}.&\numeq\cr
}
\namelasteq{\evanescent}
$$
Notice that the integrand of the integrals in eq.~{\evanescent\ formally
vanishes in the limit $d\rightarrow 4$, since it contains the evanescent
term $\hat{q}^2$. However, the limit $d\rightarrow 4$ of these integrals
although finite is not zero. Indeed, if we take into account that 
$\hat{q}^2=q^{\alpha}q^{\beta}\ghat_{\alpha\beta}$, we readily see that
what we are facing is the computation of integrals which are 
UV divergent by power-counting at $d=4$ and which will develop a 
simple pole at $d=4$ when computed in dimensional regularization (notice that
the integrals at hand are IR finite by power-counting at nonexceptional 
momenta). This pole will be canceled at the end of the day by the evanescent
(order $d-4$) contribution coming from  the contraction with 
$\ghat_{\alpha\beta}$, yielding a polynomial in the external momenta (short
distance operator)  as value for  $\triangle^{(1)}_{\mu_1\mu_2}(p_1,p_2;d)$ 
at $d=4$. We have thus explained the nonabelian chiral anomaly of 
eq.~\triangleanomaly\ as an UV effect. Indeed, a little computation 
shows that the integrals in eq.~\evanescent\ yield the following result
$$
\triangle^{(1)}_{\mu_1\mu_2}(p_1,p_2;d=4)\,=\,
-{1\over 24\pi^2}\,\varepsilon_{\mu_1\mu_2\alpha\beta}\, 
p_1^{\alpha}p_2^{\beta}.
\eqno\numeq\namelasteq\triangleoneresult
$$
The reader may find  the following integrals useful
$$
\eqalign{
&\int{d^d q\over(2\pi)^d}
{\hat{q}^2\over
q^2\, (q+p_2)^2\, (q+p_1+p_2)^2}=
-{i\over 16\pi^2}\big({1\over 2}\big)
\,+\,O(d-4),\cr
&\int{d^d q\over(2\pi)^d}
{\hat{q}^2\,\bar{q}^{\alpha}\over
q^2\, (q+p_2)^2\, (q+p_1+p_2)^2}={i\over 16\pi^2}
\big({1\over 6}\big)
(\bar{p}_1+2\bar{p}_2)^{\alpha}\,+\,O(d-4).\cr
}
$$
It is clear that for $\triangle^{(2)}_{\mu_1\mu_2}(p_1,p_2;d)$ in 
eq.~\contractedintegrals\ one will obtain the following finite answer 
$$
\triangle^{(2)}_{\mu_1\mu_2}(p_1,p_2;d=4)\,=\,
-{1\over 24\pi^2}\,\varepsilon_{\mu_1\mu_2\alpha\beta}\, 
p_1^{\alpha}p_2^{\beta}.
\eqno\numeq\namelasteq\triangletworesult
$$
Finally, if we substitute this result and eq.~\triangleoneresult\ in 
eq.~\trianglegraphs, we will recover the one-loop triangle anomaly of
eq.~\triangleanomaly .

A completely similar analysis can be done for the chiral 
theory defined by the action in eq.~\antifundamentalaction. Let us 
move on and compute   
$p_3^{\mu_3}\,\Gamma_{\mu_1\mu_2\mu_3}^{a_1 a_2 a_3}(p_1,p_2)^{\eps}$
for the theory with adjoint fermionic matter. The classical action
of this theory is given in eq.~\adjointaction. The Ward identity that
should hold if the  gauge symmetry of the classical theory is a symmetry
of the quantum theory reads
$$
\int d^4x\; \omega^{\io}_{\;\id} \star\,
\partial_{\mu}{\delta \Gamma[\A]\over \delta \A^{\io}_{\mu\,\id}}\,=\,
i\,\int d^4x\; \omega^{\io}_{\id} \star\bigg[ \A^{\id}_{\mu\,\ih}\star
            {\delta \Gamma[\A]\over \delta \A^{\io}_{\mu\,\ih}}-
{\delta \Gamma[\A]\over \delta \A^{\ih}_{\mu\,\id}}\star 
\A^{\ih}_{\mu\,\io}\bigg].
$$
For $p_3^{\mu_3}\,\Gamma_{\mu_1\mu_2\mu_3}^{a_1 a_2 a_3}(p_1,p_2)^{\eps}$,
the previous equation boils down to 
$$
p_3^{\mu_3}\,\Gamma_{\mu_1\mu_2\mu_3}^{a_1 a_2 a_3}(p_1,p_2)^{\eps}=0.
$$
To obtain this equation it is necessary to take into account that 
the two-point contribution to $\Gamma[\A]$ has no pseudotensor contribution.
 
The dimensional regularization counterpart of the action 
in eq.~\adjointaction\ will have for us the following expression
$$
{\rm S}\;=\;\intd^{d}x\; \bar{\psi}^{k}_{\;i}\star(i\prslash\psi^{i}_{\;k}
+ \A^{i}_{\mu\,j}\star \gambar^{\mu}{\rm P}_{+}\psi^{j}_{\;k}
- \gambar^{\mu}{\rm P}_{+}\psi^{i}_{\;j}\star \A^{j}_{\mu\,k}),
$$ 
with the same notation as at the beginning of this section. Instead of 
deriving Feynman rules from this action and compute 
$p_3^{\mu_3}\,\Gamma_{\mu_1\mu_2\mu_3}^{a_1 a_2 a_3}(p_1,p_2)^{\eps}$ 
from the corresponding triangle diagrams in fig. 2, we shall 
follow an alternative procedure which will supply a more thorough 
understanding of the final answer. Let us introduce first the 
following chiral current  in the ``d-dimensional'' space of dimensional
regularization 
$$
j^{a}_{\;\mu}(x)\equiv i{\delta S[\A] \over \A^a_{\mu}(x)}\equiv
j^{a\,-}_{\;\,\mu}(x)+j^{a\,+}_{\;\,\mu}(x),
\eqno\numeq\namelasteq\chiralcurrent
$$
where
$$
\eqalignno{&\qquad\qquad\qquad j^{a\,-}_{\;\,\mu}(x)=
-i\,\psi^{j}_{\;\;k\,\beta}\star\bar{\psi}^{k}_{\;\;i\,\alpha}(x)\, 
{\rm T}^{a\,i}_{\quad j}\,
\Big(\gambar^{\mu}{\rm P}_{+}\Big)_{\alpha\beta}\cr 
&{\rm and}\cr
&\qquad\qquad\qquad j^{a\,+}_{\;\,\mu}(x)=
-i\,\bar{\psi}^{k}_{\;\;i\,\alpha}\star \psi^{i}_{\;j\,\beta}(x) 
{\rm T}^{a\,j}_{\quad k}\,
\Big(\gambar^{\mu}{\rm P}_{+}\Big)_{\alpha\beta}.&\numeq\cr
}
\namelasteq{\currents}
$$
Let $j^{a\,(\cdot)}_{\;\,\mu}(p)$ be given by  
$$j^{a\,(\cdot)}_{\;\,\mu}(x)\,=\,
\int\,{d^4 p\over (4\pi)^4}\; e^{ip x}\;j^{a\,(\cdot)}_{\;\,\mu}(p).
$$ 
Then,
the three-point function (eq.~\threepointfunction) in momentum space reads
$$
\Gamma_{\mu_1\mu_2\mu_3}^{a_1 a_2 a_3}(p_1,p_2,p_3)=
\langle j^{a_1}_{\;\;\;\mu_1}(p_1)\,j^{a_2}_{\;\;\;\mu_2}(p_2)\,
j^{a_3}_{\;\;\;\mu_3}(p_3) \rangle_{\rm con}.
$$
Where the subscript ``con'' refers to the connected part of the
corresponding Green function. Throughout this paper, vacuum expectations 
values are computed with  the free fermionic action. 
Taking into account eq.~\chiralcurrent, we obtain 
$$
\Gamma_{\mu_1\mu_2\mu_3}^{a_1 a_2 a_3}(p_1,p_2,p_3)\,=\,
\Gamma_{\mu_1\mu_2\mu_3}^{a_1 a_2 a_3}(p_1,p_2,p_3)_{\rm P}\,+\,
\Gamma_{\mu_1\mu_2\mu_3}^{a_1 a_2 a_3}(p_1,p_2,p_3)_{\rm NP},
$$
where
$$
\Gamma_{\mu_1\mu_2\mu_3}^{a_1 a_2 a_3}(p_1,p_2,p_3)_{\rm P}\!=
\langle j^{a_1\,-}_{\;\;\;\mu_1}(p_1)\,j^{a_2\,-}_{\;\;\;\mu_2}(p_2)\,
j^{a_3\,-}_{\;\;\;\mu_3}(p_3) \rangle_{\rm con}\!+
\langle j^{a_1\,+}_{\;\;\;\mu_1}(p_1)\,j^{a_2\,+}_{\;\;\;\mu_2}(p_2)\,
j^{a_3\,+}_{\;\;\;\mu_3}(p_3) \rangle_{\rm con}
\eqno\numeq\namelasteq\planarcont
$$
and
$$
\eqalignno{\Gamma_{\mu_1\mu_2\mu_3}^{a_1 a_2 a_3}(p_1,p_2,p_3)_{\rm NP}\!=
&\langle j^{a_1\,-}_{\;\;\;\mu_1}(p_1)j^{a_2\,-}_{\;\;\;\mu_2}(p_2)
j^{a_3\,+}_{\;\;\;\mu_3}(p_3) \rangle_{\rm con}\!+
\langle j^{a_1\,+}_{\;\;\;\mu_1}(p_1)j^{a_2\,+}_{\;\;\;\mu_2}(p_2)
j^{a_3\,-}_{\;\;\;\mu_3}(p_3) \rangle_{\rm con}\cr
&\langle j^{a_1\,-}_{\;\;\;\mu_1}(p_1)j^{a_2\,+}_{\;\;\;\mu_2}(p_2)
j^{a_3\,-}_{\;\;\;\mu_3}(p_3) \rangle_{\rm con}\!+
\langle j^{a_1\,+}_{\;\;\;\mu_1}(p_1)j^{a_2\,-}_{\;\;\;\mu_2}(p_2)
j^{a_3\,+}_{\;\;\;\mu_3}(p_3) \rangle_{\rm con}\cr
&\langle j^{a_1\,+}_{\;\;\;\mu_1}(p_1)j^{a_2\,-}_{\;\;\;\mu_2}(p_2)
j^{a_3\,-}_{\;\;\;\mu_3}(p_3) \rangle_{\rm con}\!+
\langle j^{a_1\,-}_{\;\;\;\mu_1}(p_1)j^{a_2\,+}_{\;\;\;\mu_2}(p_2)
j^{a_3\,+}_{\;\;\;\mu_3}(p_3) \rangle_{\rm con}.\cr
&{}&\numeq\cr
}
\namelasteq{\nonplanarcont}
$$
The subscripts ``P'' and ``NP'' refer, respectively,  to the planar 
and nonplanar parts of $\Gamma_{\mu_1\mu_2\mu_3}^{a_1 a_2 a_3}(p_1,p_2,p_3)$.
The reader may easily realize that only when the three currents 
in the correlation function carry the same superscript, $-$ or $+$, 
there is no Moyal exponential carrying the loop momenta. Notice that
each correlation function of the type $\langle j\, j\, j\rangle^{\rm con}$ 
above can be interpreted as the sum of two triangle diagrams with vertices
given by the currents of the former. 

Now, taking into account eq.~\currents, it can be easily shown that the
the Green functions contributing to the r.h.s of eq.~\planarcont\ satisfy
$$
\eqalignno{&p_3^{\mu_3}\,
\langle j^{a_1\,-}_{\;\;\;\mu_1}(p_1)\,j^{a_2\,-}_{\;\;\;\mu_2}(p_2)\,
j^{a_3\,-}_{\;\;\;\mu_3}(p_3) \rangle_{\rm con}^{\rm eps}=
(2\pi)^4 \delta(p_1+p_2+p_3)\,N\,\cr
&e^{-\iteta(p_1,p_2)}\,{\rm Tr}\,\Taone\Tatwo\Tathree
\,\triangle^{(1)}_{\mu_1\mu_2}(p_1,p_2;d)+
e^{\iteta(p_1,p_2)}\,{\rm Tr}\,\Tatwo\Taone\Tathree
\,\triangle^{(2)}_{\mu_1\mu_2}(p_1,p_2;d),\cr 
&p_3^{\mu_3}\,
\langle j^{a_1\,+}_{\;\;\;\mu_1}(p_1)\,j^{a_2\,+}_{\;\;\;\mu_2}(p_2)\,
j^{a_3\,+}_{\;\;\;\mu_3}(p_3) \rangle_{\rm con}^{\rm eps}=-
(2\pi)^4 \delta(p_1+p_2+p_3)\,N\,\cr
&e^{-\iteta(p_1,p_2)}\,{\rm Tr}\,\Taone\Tatwo\Tathree
\,\triangle^{(2)}_{\mu_1\mu_2}(p_1,p_2;d)+
e^{\iteta(p_1,p_2)}\,{\rm Tr}\,\Tatwo\Taone\Tathree
\,\triangle^{(1)}_{\mu_1\mu_2}(p_1,p_2;d),\cr
&{}&\numeq\cr
}
\namelasteq{\contractoneplan}
$$
where  $\triangle^{(1)}_{\mu_1\mu_2}(p_1,p_2;d)$ and 
$\triangle^{(2)}_{\mu_1\mu_2}(p_1,p_2;d)$ are  given in 
eq.~\contractedintegrals\ and the superscript ``eps'' indicates 
that one should keep only  contributions involving the Levi-Civita symbol. 
Now, substituting eqs.~\triangleoneresult\ and~\triangletworesult\ in 
eq.~\contractoneplan, one obtains that the following equations hold at $d=4$
$$
\eqalign{&p_3^{\mu_3}\,
\langle j^{a_1\,-}_{\;\;\;\mu_1}(p_1)\,j^{a_2\,-}_{\;\;\;\mu_2}(p_2)\,
j^{a_3\,-}_{\;\;\;\mu_3}(p_3) \rangle_{\rm con}^{\rm eps}=
-(2\pi)^4 \delta(p_1+p_2+p_3)
{1\over 24\pi^2}\,\varepsilon_{\mu_1\mu_2\alpha\beta}\, 
p_1^{\alpha}p_2^{\beta}\cr
&\quad\quad\quad N\bigg(\,{\rm Tr}\,\{\Taone,\Tatwo\}\,\Tathree\cos \teta(p_1,p_2)
-i{\rm Tr}\,[\Taone,\Tatwo]\Tathree\sin\teta(p_1,p_2)\bigg),\cr
&p_3^{\mu_3}\,
\langle j^{a_1\,+}_{\;\;\;\mu_1}(p_1)\,j^{a_2\,+}_{\;\;\;\mu_2}(p_2)\,
j^{a_3\,+}_{\;\;\;\mu_3}(p_3) \rangle_{\rm con}^{\rm eps}=
(2\pi)^4 \delta(p_1+p_2+p_3)
{1\over 24\pi^2}\,\varepsilon_{\mu_1\mu_2\alpha\beta}\, 
p_1^{\alpha}p_2^{\beta}\cr
&\quad\quad\quad N\bigg(\,{\rm Tr}\,\{\Taone,\Tatwo\}\,\Tathree\cos \teta(p_1,p_2)
+i{\rm Tr}\,[\Taone,\Tatwo]\Tathree\sin\teta(p_1,p_2)\bigg).\cr
}
$$
Hence, each correlation function of currents contributing to the r.h.s of
eq.~\planarcont\ yields an anomalous term, but its sum, i.e.,
the planar part of 
$\Gamma_{\mu_1\mu_2\mu_3}^{a_1 a_2 a_3}(p_1,p_2,p_3)^{\rm eps}$, 
carries no anomaly:
$$ 
p_3^{\mu_3}\,
\Gamma_{\mu_1\mu_2\mu_3}^{a_1 a_2 a_3}(p_1,p_2,p_3)^{\rm eps}_{\rm P}\,=\,0.
$$
The reader should notice that result we have just derived can be 
understood as follows: the sum of the two triangle diagrams contributing
to $\langle j^{a_1\,-}_{\;\;\;\mu_1}(p_1)\,j^{a_2\,-}_{\;\;\;\mu_2}(p_2)\,
j^{a_3\,-}_{\;\;\;\mu_3}(p_3) \rangle_{\rm con}$ yield a chiral anomaly 
opposite to the chiral anomaly coming from the sum of the two triangle
diagrams contributing to 
$\langle j^{a_1\,+}_{\;\;\;\mu_1}(p_1)\,j^{a_2\,+}_{\;\;\;\mu_2}(p_2)\,
j^{a_3\,+}_{\;\;\;\mu_3}(p_3) \rangle_{\rm con}$, i.e., the contribution 
coming from the fermionic modes in the fundamental representation of 
$U(N)$ moving around the loop  cancels the contribution furnished by 
the fermionic modes in the anti-fundamental representation of $U(N)$ 
propagating along the loop: recall that the adjoint representation of
$U(N)$ can be understood as the product of its fundamental and anti-fundamental
representations.

Let us now show that there is no anomaly in the pseudotensor part of the
nonplanar contribution given in eq.~\nonplanarcont. Here, of course, we shall 
meet only integrals which give UV finite results at $d=4$ -since the Moyal
exponential regulate them in the UV-, but which develop, as a consequence,  
of the UV/IR connection in noncommutative field theories,  
IR divergences as one approaches the noncommutative 
IR region $\tilde{p}=0$. Let us see whether or not they carry any anomaly. 
For the first three-current correlation function on the r.h.s of 
eq.~\nonplanarcont, one obtains the following intermediate results at $d=4$
$$
\eqalignno{&p_3^{\mu_3}\,
\langle j^{a_1\,-}_{\;\;\;\mu_1}(p_1)\,j^{a_2\,-}_{\;\;\;\mu_2}(p_2)\,
j^{a_3\,+}_{\;\;\;\mu_3}(p_3) \rangle_{\rm con}^{\rm eps}=
(2\pi)^4 \delta(p_1+p_2+p_3)\Tr(\Taone\Tatwo)\Tr\, \Tathree\cr
&\qquad\qquad\qquad\Big[
e^{\iteta(p_1,p_2)}\triangle^{(1)-}_{\mu_1\mu_2}(p_1,p_2|\tilde{p}_3)+
e^{-\iteta(p_1,p_2)}\triangle^{(2)-}_{\mu_1\mu_2}(p_1,p_2|\tilde{p}_3)\big],
&\numeq\cr
}
\namelasteq{\contractednon}
$$
with
$$
\eqalign{&\triangle^{(1)-}_{\mu_1\mu_2}(p_1,p_2|\tilde{p}_3)\,=\,
\int{d^4 q\over(2\pi)^4}\,e^{-i\theta(q,p_3)}\,
{{\rm tr}^{\,\eps}\big\{ (\qslash+\poneslash)\,\gamma_{\mu_1}{\rm P}_{+}\,
\qslash\,\gamma_{\mu_2}{\rm P}_{+}\,(\qslash-\ptwoslash)(\poneslash+
\ptwoslash){\rm P}_{+}\big\}
\over (q^2+i0^+) ((q+p_1)^2+i0^+) ((q-p_2)^2+i0^+)},\cr
&{\rm and}\cr
&\triangle^{(2)-}_{\mu_1\mu_2}(p_1,p_2|\tilde{p}_3)\,=\,
\int{d^4 q\over(2\pi)^4}\,e^{-i\theta(q,p_3)}\,
{{\rm tr}^{\,\eps}\big\{ (\qslash+\ptwoslash)\,\gamma_{\mu_2}{\rm P}_{+}\,
\qslash\,\gamma_{\mu_1}{\rm P}_{+}\,(\qslash-\poneslash)(\poneslash+
\ptwoslash){\rm P}_{+}\big\}
\over (q^2+i0^+) ((q+p_2)^2+i0^+) ((q-p_1)^2+i0^+)}.\cr
} 
$$
In the previous integrals $p_1+p_2+p_3=0$. Notice the characteristic Moyal
factor, $e^{-i\theta(q,p_3)}$, of  a nonplanar contribution. The integrals
are well-defined provided we are off the noncommutative IR region defined
by $\tilde{p}_3^2=0$. Let us  show now that  
$$
e^{\iteta(p_1,p_2)}\triangle^{(1)-}_{\mu_1\mu_2}(p_1,p_2|\tilde{p}_3)+
e^{-\iteta(p_1,p_2)}\triangle^{(2)-}_{\mu_1\mu_2}(p_1,p_2|\tilde{p}_3) = 0.
\eqno\numeq\namelasteq\itvanishes
$$
If we change variables $q\rightarrow q+p_2$ and $q\rightarrow q+p_1$  in
$\triangle^{(1)-}_{\mu_1\mu_2}(p_1,p_2|\tilde{p}_3)$ and 
$\triangle^{(2)-}_{\mu_1\mu_2}(p_1,p_2|\tilde{p}_3)$, respectively,  
and use the cyclicity of the trace, we obtain 
$$
\eqalignno{&e^{\iteta(p_1,p_2)}
\triangle^{(1)-}_{\mu_1\mu_2}(p_1,p_2|\tilde{p}_3)+
e^{-\iteta(p_1,p_2)}\triangle^{(2)-}_{\mu_1\mu_2}(p_1,p_2|\tilde{p}_3) =\cr
&e^{-\iteta(p_1,p_2)}
\int{d^4 q\over(2\pi)^4}\,e^{-i\theta(q,p_3)}\,
{\rm tr}^{\,\eps}\big\{ {1\over \qslash}(\poneslash+\ptwoslash){\rm P}_{+}\,
{1\over \qslash +\poneslash+\ptwoslash}\,\gamma_{\mu_1}{\rm P}_{+}\,
{1\over \qslash+\ptwoslash}\,\gamma_{\mu_2}{\rm P}_{+}\big\}+\cr
&e^{\iteta(p_1,p_2)}
\int{d^4 q\over(2\pi)^4}\,e^{-i\theta(q,p_3)}\,
{\rm tr}^{\,\eps}\big\{ {1\over \qslash}(\poneslash+\ptwoslash){\rm P}_{+}\,
{1\over \qslash +\poneslash+\ptwoslash}\,\gamma_{\mu_2}{\rm P}_{+}\,
{1\over \qslash+\poneslash}\,\gamma_{\mu_1}{\rm P}_{+}\big\}.\cr
&{}&\numeq\cr
} 
\namelasteq{\zeroint}
$$
Now, using the equations $\gamma_{\mu}\gamma_{\nu}{\rm P}_{+}={\rm P}_{+}
\gamma_{\mu}\gamma_{\nu}$,  ${\rm P}_{+}^2={\rm P}_{+}$ and
$$
(\poneslash+\ptwoslash)\gamma_5=-\qslash\gamma_5-
\gamma_5(\qslash+\poneslash+\ptwoslash),
$$
one readily casts the r.h.s of eq.~\zeroint\ into the form
$$
\eqalignno{
-{1\over 2}\,e^{-\iteta(p_1,p_2)}
\int{d^4 q\over(2\pi)^4}\,e^{-i\theta(q,p_3)}\,\bigg[&
{\rm tr}\big\{ \gamma_5\,
{1\over \qslash +\poneslash+\ptwoslash}\,\gamma_{\mu_1}\,
{1\over \qslash+\ptwoslash}\,\gamma_{\mu_2}\big\}+\cr
&{\rm tr}\big\{
{1\over \qslash }\,\gamma_5\gamma_{\mu_1}\,
{1\over \qslash+\ptwoslash}\,\gamma_{\mu_2}\big\}\bigg]\cr
-{1\over 2}e^{\iteta(p_1,p_2)}
\int{d^4 q\over(2\pi)^4}\,e^{-i\theta(q,p_3)}\,\bigg[&
{\rm tr}\big\{\gamma_5\,
{1\over \qslash +\poneslash+\ptwoslash}\,\gamma_{\mu_2}\,
{1\over \qslash+\poneslash}\,\gamma_{\mu_1}\big\}+\cr
&{\rm tr}\big\{
{1\over \qslash }\,\gamma_5\gamma_{\mu_2}\,
{1\over \qslash+\poneslash}\,\gamma_{\mu_1}\big\}\bigg].&\numeq\cr
} 
\namelasteq{\zerointuno}
$$
Some Dirac algebra leads, respectively, to the following expressions
$$ 
\eqalignno{&
{\rm tr}\big\{ \gamma_5\,
{1\over \qslash +\poneslash+\ptwoslash}\,\gamma_{\mu_1}\,
{1\over \qslash+\ptwoslash}\,\gamma_{\mu_2}\big\}=
-{\rm tr}\big\{ 
{1\over \qslash+\ptwoslash}\,\gamma_5\gamma_{\mu_2}\,
{1\over \qslash +\poneslash+\ptwoslash}\,\gamma_{\mu_1}\big\},\cr
&{\rm tr}\big\{\gamma_5 
{1\over \qslash +\poneslash+\ptwoslash}\,\gamma_{\mu_2}\,
{1\over \qslash+\poneslash}\,\gamma_{\mu_1}\big\}=
-{\rm tr}\big\{{1\over \qslash+\poneslash}\,\gamma_5\gamma_{\mu_1}
{1\over \qslash +\poneslash+\ptwoslash}\,\gamma_{\mu_2}\}.
&\numeq\cr
} 
\namelasteq{\algebraic}
$$
Substituting these equations in eq.~\zerointuno\ and performing appropriate
momentum shifts, one easily shows that, in eq.~\zerointuno, the first integral 
cancels the fourth integral and the second integral cancels the third one:
thus proving that the eq.~\itvanishes\ actually holds. We get finally
$$
p_3^{\mu_3}\,
\langle j^{a_1\,-}_{\;\;\;\mu_1}(p_1)\,j^{a_2\,-}_{\;\;\;\mu_2}(p_2)\,
j^{a_3\,+}_{\;\;\;\mu_3}(p_3) \rangle_{\rm con}^{\rm eps}=0;
\eqno\numeq\namelasteq\zerofirst
$$
a result which is obtained by substituting eq.~\itvanishes\ in 
eq.~\contractednon. The same conclusion can be reached, using 
completely analogous methods, for the  three-current correlation
function $\langle j^{a_1\,+}_{\;\;\;\mu_1}(p_1)j^{a_2\,+}_{\;\;\;\mu_2}(p_2)
j^{a_3\,-}_{\;\;\;\mu_3}(p_3) \rangle_{\rm con}$:
$$
p_3^{\mu_3}\,
\langle j^{a_1\,+}_{\;\;\;\mu_1}(p_1)\,j^{a_2\,+}_{\;\;\;\mu_2}(p_2)\,
j^{a_3\,-}_{\;\;\;\mu_3}(p_3) \rangle_{\rm con}^{\rm eps}=0.
\eqno\numeq\namelasteq\zerosecond
$$
Things do not work the same way for the remaining Green functions on the 
r.h.s. of eq.~\nonplanarcont. Actually, each three-current correlation
function  gives a contribution, vanishing  the sum of them all.
Let us see it.  Some algebra leads to 
$$
\eqalignno{
&p_3^{\mu_3}\,
\langle j^{a_1\,-}_{\;\;\;\mu_1}(p_1)\,j^{a_2\,+}_{\;\;\;\mu_2}(p_2)\,
j^{a_3\,-}_{\;\;\;\mu_3}(p_3) \rangle_{\rm con}^{\rm eps}=
(2\pi)^4 \delta(p_1+p_2+p_3)\Tr(\Taone\Tathree)\Tr\, \Tatwo\cr
&\qquad\qquad\qquad\Big[
e^{-\iteta(p_1,p_2)}\triangle^{(1)-}_{\mu_1\mu_2}(p_1,p_2|\tilde{p}_2)+
e^{\iteta(p_1,p_2)}\triangle^{(2)-}_{\mu_1\mu_2}(p_1,p_2|\tilde{p}_2)
\big],\cr
&p_3^{\mu_3}\,
\langle j^{a_1\,+}_{\;\;\;\mu_1}(p_1)\,j^{a_2\,-}_{\;\;\;\mu_2}(p_2)\,
j^{a_3\,+}_{\;\;\;\mu_3}(p_3) \rangle_{\rm con}^{\rm eps}=-
(2\pi)^4 \delta(p_1+p_2+p_3)\Tr(\Taone\Tathree)\Tr\, \Tatwo\cr
&\qquad\qquad\qquad\Big[
e^{\iteta(p_1,p_2)}\triangle^{(1)+}_{\mu_1\mu_2}(p_1,p_2|\tilde{p}_2)+
e^{-\iteta(p_1,p_2)}\triangle^{(2)+}_{\mu_1\mu_2}(p_1,p_2|\tilde{p}_2)
\big],\cr
&p_3^{\mu_3}\,
\langle j^{a_1\,+}_{\;\;\;\mu_1}(p_1)\,j^{a_2\,-}_{\;\;\;\mu_2}(p_2)\,
j^{a_3\,-}_{\;\;\;\mu_3}(p_3) \rangle_{\rm con}^{\rm eps}=
(2\pi)^4 \delta(p_1+p_2+p_3)\Tr(\Tatwo\Tathree)\Tr\, \Taone\cr
&\qquad\qquad\qquad\Big[
e^{-\iteta(p_1,p_2)}\triangle^{(1)-}_{\mu_1\mu_2}(p_1,p_2|\tilde{p}_1)+
e^{\iteta(p_1,p_2)}\triangle^{(2)-}_{\mu_1\mu_2}(p_1,p_2|\tilde{p}_1)\big],\cr
&p_3^{\mu_3}\,
\langle j^{a_1\,-}_{\;\;\;\mu_1}(p_1)\,j^{a_2\,+}_{\;\;\;\mu_2}(p_2)\,
j^{a_3\,+}_{\;\;\;\mu_3}(p_3) \rangle_{\rm con}^{\rm eps}=-
(2\pi)^4 \delta(p_1+p_2+p_3)\Tr(\Tatwo\Tathree)\Tr\, \Taone\cr
&\qquad\qquad\qquad\Big[
e^{\iteta(p_1,p_2)}\triangle^{(1)+}_{\mu_1\mu_2}(p_1,p_2|\tilde{p}_1)+
e^{-\iteta(p_1,p_2)}\triangle^{(2)+}_{\mu_1\mu_2}(p_1,p_2|\tilde{p}_1)\big].
&\numeq\cr
}
\namelasteq{\contractednonrem}
$$
In this equation, the contributions denoted by
$\triangle^{(1)\pm}_{\mu_1\mu_2}(p_1,p_2|\tilde{p}_i)$ and 
$ \triangle^{(2)\pm}_{\mu_1\mu_2}(p_1,p_2|\tilde{p}_i)$, with 
$i=1$ and $2$, are given by the
following integrals
$$
\eqalign{&\triangle^{(1)\pm}_{\mu_1\mu_2}(p_1,p_2|\tilde{p}_i)\,=\,
\int{d^4 q\over(2\pi)^4}\,e^{\pm i\theta(q,p_i)}\,
{{\rm tr}^{\,\eps}\big\{ (\qslash+\poneslash)\,\gamma_{\mu_1}{\rm P}_{+}\,
\qslash\,\gamma_{\mu_2}{\rm P}_{+}\,(\qslash-\ptwoslash)(\poneslash+
\ptwoslash){\rm P}_{+}\big\}
\over (q^2+i0^+) ((q+p_1)^2+i0^+) ((q-p_2)^2+i0^+)},\cr
&{\rm and}\cr
&\triangle^{(2)\pm}_{\mu_1\mu_2}(p_1,p_2\tilde{p}_i)\,=\,
\int{d^4 q\over(2\pi)^4}\,e^{\pm i\theta(q,p_i)}\,
{{\rm tr}^{\,\eps}\big\{ (\qslash+\ptwoslash)\,\gamma_{\mu_2}{\rm P}_{+}\,
\qslash\,\gamma_{\mu_1}{\rm P}_{+}\,(\qslash-\poneslash)(\poneslash+
\ptwoslash){\rm P}_{+}\big\}
\over (q^2+i0^+) ((q+p_2)^2+i0^+) ((q-p_1)^2+i0^+)}.\cr
} 
$$
Using the same variety of tricks that led to eq.~\itvanishes, one shows that
now
$$
\eqalignno{
&e^{\mp\iteta(p_1,p_2)}\triangle^{(1)\mp}_{\mu_1\mu_2}(p_1,p_2|\tilde{p}_i)+
e^{\pm\iteta(p_1,p_2)}\triangle^{(2)\mp}_{\mu_1\mu_2}(p_1,p_2|\tilde{p}_i)=\cr
&\mp 4\sin{1\over 2}\theta(p_1,p_2)\;
\varepsilon_{\mu_1\mu_2\alpha\beta}\, \int{d^4 q\over(2\pi)^4}\,
e^{\mp i\theta(q,p_i)}\;
{q^{\alpha}p^{\beta}_i\over (q^2+i0^+)((q+p_i)^2+i0^+)},
&\numeq\cr
}
\namelasteq{\itisnotzero}
$$
where $i=1$ and $2$. For the sake of the reader, we shall spell out the 
computations leading to the previous equation.
Let us change variables $q\rightarrow q+p_2$ and $q\rightarrow q+p_1$  in
$\triangle^{(1)\mp}_{\mu_1\mu_2}(p_1,p_2|\tilde{p}_2)$ and 
$\triangle^{(2)\mp}_{\mu_1\mu_2}(p_1,p_2|\tilde{p}_2)$, respectively,  
and use the cyclicity of the trace, to obtain 
$$
\eqalignno{&e^{\mp\iteta(p_1,p_2)}
\triangle^{(1)\mp}_{\mu_1\mu_2}(p_1,p_2|\tilde{p}_2)+
e^{\pm\iteta(p_1,p_2)}\triangle^{(2)\mp}_{\mu_1\mu_2}(p_1,p_2|\tilde{p}_2) =\cr
&e^{\mp\iteta(p_1,p_2)}
\int{d^4 q\over(2\pi)^4}\,e^{\mp i\theta(q,p_2)}\,
{\rm tr}^{\,\eps}\big\{ {1\over \qslash}(\poneslash+\ptwoslash){\rm P}_{+}\,
{1\over \qslash +\poneslash+\ptwoslash}\,\gamma_{\mu_1}{\rm P}_{+}\,
{1\over \qslash+\ptwoslash}\,\gamma_{\mu_2}{\rm P}_{+}\big\}+\cr
&e^{\mp\iteta(p_1,p_2)}
\int{d^4 q\over(2\pi)^4}\,e^{\mp i\theta(q,p_2)}\,
{\rm tr}^{\,\eps}\big\{ {1\over \qslash}(\poneslash+\ptwoslash){\rm P}_{+}\,
{1\over \qslash +\poneslash+\ptwoslash}\,\gamma_{\mu_2}{\rm P}_{+}\,
{1\over \qslash+\poneslash}\,\gamma_{\mu_1}{\rm P}_{+}\big\}.\cr
&{}&\numeq\cr
} 
\namelasteq{\nonflipped}
$$
Notice that, unlike eq.~\zeroint, the exponential factor in front of
each integral is the same. This will turn out to be of the utmost 
importance. Next, let us use the equations 
$\gamma_{\mu}\gamma_{\nu}{\rm P}_{+}={\rm P}_{+}\gamma_{\mu}\gamma_{\nu}$,  
${\rm P}_{+}^2={\rm P}_{+}$ and
$$
(\poneslash+\ptwoslash)\gamma_5=-\qslash\gamma_5-
\gamma_5(\qslash+\poneslash+\ptwoslash),
$$
to cast the r.h.s of eq.~\nonflipped\ into the form
$$
\eqalignno{
-{1\over 2}\,e^{\mp\iteta(p_1,p_2)}
\int{d^4 q\over(2\pi)^4}\,e^{\mp i\theta(q,p_2)}\,\bigg[&
{\rm tr}\big\{ \gamma_5\,
{1\over \qslash +\poneslash+\ptwoslash}\,\gamma_{\mu_1}\,
{1\over \qslash+\ptwoslash}\,\gamma_{\mu_2}\big\}+\cr
&{\rm tr}\big\{
{1\over \qslash }\,\gamma_5\gamma_{\mu_1}\,
{1\over \qslash+\ptwoslash}\,\gamma_{\mu_2}\big\}\bigg]\cr
-{1\over 2}e^{\mp\iteta(p_1,p_2)}
\int{d^4 q\over(2\pi)^4}\,e^{\mp i\theta(q,p_2)}\,\bigg[&
{\rm tr}\big\{\gamma_5\,
{1\over \qslash +\poneslash+\ptwoslash}\,\gamma_{\mu_2}\,
{1\over \qslash+\poneslash}\,\gamma_{\mu_1}\big\}+\cr
&{\rm tr}\big\{
{1\over \qslash }\,\gamma_5\gamma_{\mu_2}\,
{1\over \qslash+\poneslash}\,\gamma_{\mu_1}\big\}\bigg].&\numeq\cr
} 
\namelasteq{\nonflippeduno}
$$
Recall that ${\rm tr}^{\,\eps}$ means that one only keeps contributions that
carry the Levi-Civita symbol. Taking into account eq.~\algebraic, one
obtains that eq.~\nonflippeduno\ can be written as  follows
$$
\eqalignno{
-{1\over 2}\,e^{\mp\iteta(p_1,p_2)}
\int{d^4 q\over(2\pi)^4}\,e^{\mp i\theta(q,p_2)}\,\bigg[-&
{\rm tr}\big\{
{1\over \qslash +\ptwoslash}\,\gamma_5\,\gamma_{\mu_2}\,
{1\over \qslash+\poneslash+\ptwoslash}\,\gamma_{\mu_1}\big\}\cr
+&{\rm tr}\big\{
{1\over \qslash }\,\gamma_5\gamma_{\mu_1}\,
{1\over \qslash+\ptwoslash}\,\gamma_{\mu_2}\big\}\bigg]\cr
-{1\over 2}e^{\mp\iteta(p_1,p_2)}
\int{d^4 q\over(2\pi)^4}\,e^{\mp i\theta(q,p_2)}\,\bigg[-&
{\rm tr}\big\{
{1\over \qslash +\poneslash}\,\gamma_5\,\gamma_{\mu_1}\,
{1\over \qslash+\poneslash+\ptwoslash}\,\gamma_{\mu_2}\big\}\cr
+&{\rm tr}\big\{
{1\over \qslash }\,\gamma_5\gamma_{\mu_2}\,
{1\over \qslash+\poneslash}\,\gamma_{\mu_1}\big\}\bigg].&\numeq\cr
} 
\namelasteq{\nonflippeddos}
$$
Let us next make the following shifts, $q\rightarrow q-p_2$ and 
$q\rightarrow q-p_1$, in the first and third integrals in eq.~\nonflippeddos.
Then, we readily see that the first integral cancels the fourth    
integral of eq.~\nonflippeddos, but the sum of the second and 
third integrals of eq.~\nonflippeddos\ yield 
$$
{1\over 2}\,\bigg(e^{\pm\iteta(p_1,p_2)}-e^{\mp\iteta(p_1,p_2)}\bigg)
\int{d^4 q\over(2\pi)^4}\,e^{\mp i\theta(q,p_2)}\,{\rm tr}\big\{
{1\over \qslash }\,\gamma_5\gamma_{\mu_1}\,
{1\over \qslash+\ptwoslash}\,\gamma_{\mu_2}\big\}.
$$
From this equation one obtains eq.~\itisnotzero\ for $i=2$.
Let us now  replace  the integral in eq.~\itisnotzero\ with its value, 
which can be found in the appendix.  One   obtains, for $i=2$,  that 
$$
\eqalignno{
&e^{\mp\iteta(p_1,p_2)}\triangle^{(1)\mp}_{\mu_1\mu_2}(p_1,p_2|\tilde{p}_i)+
e^{\pm\iteta(p_1,p_2)}\triangle^{(2)\mp}_{\mu_1\mu_2}(p_1,p_2|\tilde{p}_i)=
 {1\over 2\pi^2}\sin{1\over 2}\theta(p_1,p_2)\cr
&\varepsilon_{\mu_1\mu_2\alpha\beta}\, {\tilde{p}_i^{\alpha}p_i^{\beta}
\over \tilde{p}_i^2}
\,\int_{0}^{1}\,dx\,\sqrt{\tilde{p}_i^2(-p_i^2-i0^+)x(1-x)}\;\;
{\rm K}_1\bigg(\sqrt{\tilde{p}_i^2(-p_i^2-i0^+)x(1-x)}\bigg);
&\numeq\cr
}
\namelasteq{\itisnonzero}
$$
a result which is also valid for $i=1$. Let us warn the reader that we use the 
notation $\tilde{p}_i^{\mu}=\theta^{\mu\nu}p_{i\,\nu}$ and
$\tilde{p}^2_i\equiv p_{i\,\mu}\,\theta^{\mu\rho}
\eta_{\rho\sigma}\theta^{\sigma\nu}\,p_{i\,\nu}$, so that
$\tilde{p}^2_i\geq 0$.
Substituting this result in eqs.~\contractednonrem, one comes to the 
conclusion that there is a pairwise cancellation mechanism at work:
$$
\eqalignno{
&p_3^{\mu_3}\,
\langle j^{a_1\,-}_{\;\;\;\mu_1}(p_1)\,j^{a_2\,+}_{\;\;\;\mu_2}(p_2)\,
j^{a_3\,-}_{\;\;\;\mu_3}(p_3) \rangle_{\rm con}^{\rm eps}+
p_3^{\mu_3}\,
\langle j^{a_1\,+}_{\;\;\;\mu_1}(p_1)\,j^{a_2\,-}_{\;\;\;\mu_2}(p_2)\,
j^{a_3\,+}_{\;\;\;\mu_3}(p_3) \rangle_{\rm con}^{\rm eps}=0,\cr
&p_3^{\mu_3}\,
\langle j^{a_1\,+}_{\;\;\;\mu_1}(p_1)\,j^{a_2\,-}_{\;\;\;\mu_2}(p_2)\,
j^{a_3\,-}_{\;\;\;\mu_3}(p_3) \rangle_{\rm con}^{\rm eps}+
p_3^{\mu_3}\,
\langle j^{a_1\,-}_{\;\;\;\mu_1}(p_1)\,j^{a_2\,+}_{\;\;\;\mu_2}(p_2)\,
j^{a_3\,+}_{\;\;\;\mu_3}(p_3) \rangle_{\rm con}^{\rm eps}=0.
&\numeq\cr
}
\namelasteq{\remainingzeroes}
$$
Finally, taking into account eqs.~\nonplanarcont,~\zerofirst,~\zerosecond\ and~
\remainingzeroes, one concludes that in the pseudotensor part of 
$\Gamma^{a_1 a_2 a_3}_{\mu_1 \mu_2 \mu_3}(p_1,p_2,p_3)_{\rm NP}$ 
no chiral gauge anomaly occurs, i.e.,
$$ 
p_3^{\mu_3}\,
\Gamma_{\mu_1\mu_2\mu_3}^{a_1 a_2 a_3}(p_1,p_2,p_3)^{\rm eps}_{\rm NP}=0.
$$
We have thus shown that a noncommutative $U(N)$ chiral theory with only  
chiral adjoint fermions do not present a chiral anomaly in the three point 
function (triangle anomaly). The descent equations~\cite{\Bonora} leads to
the conclusion that noncommutative $U(N)$ chiral gauge theory with only
adjoint fermions is anomaly free. 

Let us now study the gauge anomalies of the theory with  action in 
eq.~\biaction. If this theory were gauge invariant at the quantum level
the Ward identities should read thus
$$
\eqalignno{&\int d^4x\; \omega^{\io}_{\;\id} \star\,
\partial_{\mu}{\delta \Gamma[\A,\B]\over \delta \A^{\io}_{\mu\,\id}}\,=\,
i\,\int d^4x\; \omega^{\io}_{\;\id} \star\bigg[ \A^{\id}_{\mu\,\ih}\star
            {\delta \Gamma[\A,\B]\over \delta \A^{\io}_{\mu\,\ih}}-
{\delta \Gamma[\A,\B]\over \delta \A^{\ih}_{\mu\,\id}}
\star \A^{\ih}_{\mu\,\io}\bigg],
\cr 
&\int d^4x\; \chi^{\jo}_{\;\,\jd} \star\,
\partial_{\mu}{\delta \Gamma[\A,\B]\over \delta \B^{\jo}_{\mu\,\jd}}\,=\,
i\,\int d^4x\; \chi^{\jo}_{\;\,\jd} \star\bigg[ \B^{\jd}_{\mu\,\jh}\star
            {\delta \Gamma[\A,\B]\over \delta \B^{\jo}_{\mu\,\jh}}-
{\delta \Gamma[\A,\B]\over \delta \B^{\jh}_{\mu\,\jd}}
\star \B^{\jh}_{\mu\,\jo}\bigg].\cr
&{}&\numeq\cr
}
\namelasteq{\Wardbiad}
$$
Let us introduce the following currents 
$$
j^{a}_{\;\,\mu}(x)\equiv i{\delta S[\A, \B] \over \A^{a}_{\mu}(x)}
\quad{\rm and }\quad
j^{b}_{\;\,\mu}(x)\equiv i{\delta S[\A, \B] \over \B^{b}_{\mu}(x)}.
$$
Hence,
$$
j^{a}_{\;\,\mu}(x)=
-i\,\psi^{j}_{\;\;k\,\beta}\star\bar{\psi}^{k}_{\;\;i\,\alpha}(x)\, 
{\rm T}^{a\,i}_{U(N)\; j}\,
\Big(\gambar^{\mu}{\rm P}_{+}\Big)_{\alpha\beta}
\eqno\numeq\namelasteq\Acurrent
$$
and
$$
j^{b}_{\;\,\mu}(x)=
-i\,\bar{\psi}^{k}_{\;\;i\,\alpha}\star \psi^{i}_{\;j\,\beta}(x) 
{\rm T}^{b\,j}_{U(M)\; k}\,
\Big(\gambar^{\mu}{\rm P}_{+}\Big)_{\alpha\beta},
\eqno\numeq\namelasteq\Bcurrent
$$
where ${\rm T}^a_{U(N)}$ and ${\rm T}^b_{U(M)}$ are the generators 
of $U(N)$ and $U(M)$ in the fundamental representation, respectively.
We shall also need  the following nonsinglet currents,
$$
j^{(\A)\,\io}_{\;\,\mu\;\; \id}(x)=-i\,
\psi^{\io}_{\;\;j\,\beta}\star\bar{\psi}^{j}_{\;\;\id\,\alpha}(x)
\Big(\gambar^{\mu}{\rm P}_{+}\Big)_{\alpha\beta}
\eqno\numeq\namelasteq\Anonsinglet
$$
and
$$
j^{(\B)\,\jo}_{\;\,\mu\;\; \jd}(x)=-i\,
\bar{\psi}^{\jo}_{\;\;i\,\alpha}\star \psi^{i}_{\;\jd\,\beta}(x)
\Big(\gambar^{\mu}{\rm P}_{+}\Big)_{\alpha\beta},
\eqno\numeq\namelasteq\Bnonsinglet
$$
to express the r.h.s of eq.~\Wardbiad\ in terms of correlation functions
of currents. Unlike the theories previously studied, now, 
there are nonvanishing pseudotensor contributions to the two-point 
part of $\Gamma[\A,\B]$. These contributions enter the Ward identities 
in eq.~\Wardbiad.

We have now the following independent three-current correlation functions
$$
\eqalignno{&\langle j^{a_1}_{\;\;\;\mu_1}(p_1)\,j^{a_2}_{\;\;\;\mu_2}(p_2)\,
j^{a_3}_{\;\;\;\mu_3}(p_3) \rangle_{\rm con},\;
\langle j^{b_1}_{\;\;\;\mu_1}(p_1)\,j^{b_2}_{\;\;\;\mu_2}(p_2)\,
j^{b_3}_{\;\;\;\mu_3}(p_3) \rangle_{\rm con},\cr
&\langle j^{b_1}_{\;\;\;\mu_1}(p_1)\,j^{b_2}_{\;\;\;\mu_2}(p_2)\,
j^{a_3}_{\;\;\;\mu_3}(p_3) \rangle_{\rm con},\;
\langle j^{a_1}_{\;\;\;\mu_1}(p_1)\,j^{a_2}_{\;\;\;\mu_2}(p_2)\,
j^{b_3}_{\;\;\;\mu_3}(p_3) \rangle_{\rm con},\cr
&\langle j^{a_1}_{\;\;\;\mu_1}(p_1)\,j^{b_2}_{\;\;\;\mu_2}(p_2)\,
j^{a_3}_{\;\;\;\mu_3}(p_3) \rangle_{\rm con}\quad{\rm and}\quad
\langle j^{b_1}_{\;\;\;\mu_1}(p_1)\,j^{a_2}_{\;\;\;\mu_2}(p_2)\,
j^{b_3}_{\;\;\;\mu_3}(p_3) \rangle_{\rm con}.&\numeq\cr
}
\namelasteq{\currentcor}
$$
The reader should bear in mind that the 
indices $a_i$, $i=1,\,2$ and $3$, label currents of the type defined 
in eq.~\Acurrent, whereas if a current is of the type given in 
eq.~\Bcurrent\ it carries an index $b_i$, $i=1,\,2$ and $3$.
In eq.~\currentcor\ the first two correlation functions are sums of 
only planar triangle diagrams and the last four are sums of only nonplanar 
triangle diagrams. 
That there be no breaking of the classical gauge symmetry of 
the theory at hand in the triangle diagrams, demands that 
the following equations hold
$$
\eqalignno{p_3^{\mu_3}\,
\langle j^{a_1}_{\;\;\;\mu_1}(p_1)\,j^{a_2}_{\;\;\;\mu_2}(p_2)\,
j^{a_3}_{\;\;\;\mu_3}(p_3) \rangle_{\rm con}^{\rm eps}=&\,0,\cr
p_3^{\mu_3}\,
\langle j^{b_1}_{\;\;\;\mu_1}(p_1)\,j^{b_2}_{\;\;\;\mu_2}(p_2)\,
j^{b_3}_{\;\;\;\mu_3}(p_3) \rangle_{\rm con}^{\rm eps}=&\,0,\cr
p_3^{\mu_3}\,
\langle j^{b_1}_{\;\;\;\mu_1}(p_1)\,j^{b_2}_{\;\;\;\mu_2}(p_2)\,
j^{a_3}_{\;\;\;\mu_3}(p_3) \rangle_{\rm con}^{\rm eps}=&\,0,\cr
p_3^{\mu_3}\,
\langle j^{a_1}_{\;\;\;\mu_1}(p_1)\,j^{a_2}_{\;\;\;\mu_2}(p_2)\,
j^{b_3}_{\;\;\;\mu_3}(p_3) \rangle_{\rm con}^{\rm eps}=&\,0,\cr
p_3^{\mu_3}\,
\langle j^{a_1}_{\;\;\;\mu_1}(p_1)\,j^{b_2}_{\;\;\;\mu_2}(p_2)\,
j^{a_3}_{\;\;\;\mu_3}(p_3) \rangle_{\rm con}^{\rm eps}=&-
e^{\iteta(p_1,p_2)}\,
\T^{a_1\,\io}_{\quad\ih}\,\T^{a_3\, \ih}_{\quad\id}\,
\langle j^{(\A)\,\id}_{\;\;\;\mu_1\,\io}(-p_2)\,j^{b_2}_{\;\;\;\mu_2}(p_2)\,
\rangle_{\rm con}^{\rm eps}\cr
&+e^{-\iteta(p_1,p_2)}
\T^{a_3\, \io}_{\quad\ih}\,\T^{a_1\,\ih}_{\quad\id}\,
\langle j^{(\A)\,\id}_{\;\;\;\mu_1\,\io}(-p_2)\,j^{b_2}_{\;\;\;\mu_2}(p_2)\,
\rangle_{\rm con}^{\rm eps},\cr
p_3^{\mu_3}\,
\langle j^{b_1}_{\;\;\;\mu_1}(p_1)\,j^{a_2}_{\;\;\;\mu_2}(p_2)\,
j^{b_3}_{\;\;\;\mu_3}(p_3) \rangle_{\rm con}^{\rm eps}=&-
e^{\iteta(p_1,p_2)}\,
\T^{b_1\,\jo}_{\quad\jh}\,\T^{b_3\, \jh}_{\quad\jd}\,
\langle j^{(\B)\,\jd}_{\;\;\;\mu_1\,\jo}(-p_2)\,j^{a_2}_{\;\;\;\mu_2}(p_2)\,
\rangle_{\rm con}^{\rm eps}\cr
&+e^{-\iteta(p_1,p_2)}
\T^{b_3\, \jo}_{\quad\jh}\,\T^{b_1\,\jh}_{\quad\jd}\,
\langle j^{(\B)\,\jd}_{\;\;\;\mu_1\,\jo}(-p_2)\,j^{a_2}_{\;\;\;\mu_2}(p_2)\,
\rangle_{\rm con}^{\rm eps}.\cr
&{}&\numeq\cr
}
\namelasteq{\bianomalyfreedom}
$$
Where $p_3=\!-p_1\!-p_2$. The nonsinglet currents $j^{(\A)}_{\;\,\mu}$ and
$j^{(\B)}_{\;\,\mu}$ are defined in eqs.~\Anonsinglet\ and~\Bnonsinglet,   
respectively. To obtain  the previous equation, we have taken into account 
eq.~\Wardbiad\  and the result that the only two-point 
contribution to $\Gamma[\A, \B]$ which carries a pseudontensor 
contribution is of the type
$$
\int d^4 x\,\int d^4 y\; \Tr\, \A_{\mu_1}(x)\;\Tr\, \B_{\mu_2}(y)\;
{\rm f}^{\mu_1\mu_2}\,(x, y|\theta),
$$
with
$$
\eqalign{
{\rm f}^{\mu_1\mu_2}\,(x,y|\theta)&=\int {d^4 p\over (2\pi)^4}\,
e^{-ip(x-y)}\;{\rm f}^{\mu_1\mu_2}\,(p|\tilde{p}),\cr
{\rm f}^{\mu_1\mu_2}\,(p|\tilde{p})&={i\over 4\pi^2}\,
\varepsilon_{\mu_1\mu_2\alpha\beta}\, {\tilde{p}^{\alpha}p^{\beta}
\over \tilde{p}^2}\cr
&\int_{0}^{1}\,dx\,\sqrt{\tilde{p}^2(-p^2-i0^+)x(1-x)}\;\;
{\rm K}_1\bigg(\sqrt{\tilde{p}^2(-p^2-i0^+)x(1-x)}\bigg).\cr
}
$$
This pseudotensor contribution is nonplanar and causes no anomaly.

Let us see that the first two identities in eq.~\bianomalyfreedom\ 
do not hold, so that they are anomalous,  but the all the others do.  
The computations we have carried out for the theory with adjoint fermion
fields can be readily adapted to the case at hand to obtain
$$
\eqalignno{&p_3^{\mu_3}\,
\langle j^{a_1}_{\;\;\;\mu_1}(p_1)\,j^{a_2}_{\;\;\;\mu_2}(p_2)\,
j^{a_3}_{\;\;\;\mu_3}(p_3) \rangle_{\rm con}^{\rm eps}=
-(2\pi)^4 \delta(p_1+p_2+p_3)
{1\over 24\pi^2}\,\varepsilon_{\mu_1\mu_2\alpha\beta}\, 
p_1^{\alpha}p_2^{\beta}\cr
& M\bigg(\,{\rm Tr}\,
\{\Taone_{U(N)},\Tatwo_{U(N)}\}\,\Tathree_{U(N)}\cos \teta(p_1,p_2)
-i{\rm Tr}\,[\Taone_{U(N)},\Tatwo_{U(N)}]\Tathree_{U(N)}
\sin\teta(p_1,p_2)\bigg),\cr
&p_3^{\mu_3}\,
\langle j^{b_1}_{\;\;\;\mu_1}(p_1)\,j^{b_2}_{\;\;\;\mu_2}(p_2)\,
j^{b_3}_{\;\;\;\mu_3}(p_3) \rangle_{\rm con}^{\rm eps}=
(2\pi)^4 \delta(p_1+p_2+p_3)
{1\over 24\pi^2}\,\varepsilon_{\mu_1\mu_2\alpha\beta}\, 
p_1^{\alpha}p_2^{\beta}\cr
& N\bigg(\,{\rm Tr}\,
\{\T^{b_1}_{U(M)}, \T^{b_2}_{U(M)}\}\, \T^{b_3}_{U(M)} \cos\teta(p_1,p_2)
- i{\rm Tr}\,[\T^{b_1}_{U(M)},\T^{b_2}_{U(M)}]\T^{b_3}_{U(M)}
\sin\teta(p_1,p_2)\bigg)\cr
&{}&\numeq\cr
}
\namelasteq{\anomalousplan}
$$
and 
$$
\eqalignno{
&p_3^{\mu_3}\,
\langle j^{b_1}_{\;\;\;\mu_1}(p_1)\,j^{b_2}_{\;\;\;\mu_2}(p_2)\,
j^{a_3}_{\;\;\;\mu_3}(p_3) \rangle_{\rm con}^{\rm eps}=-
(2\pi)^4 \delta(p_1+p_2+p_3)\Tr(\T^{b_1}_{U(M)}\T^{b_2}_{U(M)})\Tr\, 
\Tathree_{U(N)}\cr
&\qquad\qquad\qquad\Big[
e^{-\iteta(p_1,p_2)}\triangle^{(1)+}_{\mu_1\mu_2}(p_1,p_2|\tilde{p}_3)+
e^{\iteta(p_1,p_2)}\triangle^{(2)+}_{\mu_1\mu_2}(p_1,p_2|\tilde{p}_3)
\Big],\cr
&p_3^{\mu_3}\,
\langle j^{a_1}_{\;\;\;\mu_1}(p_1)\,j^{a_2}_{\;\;\;\mu_2}(p_2)\,
j^{b_3}_{\;\;\;\mu_3}(p_3) \rangle_{\rm con}^{\rm eps}=
(2\pi)^4 \delta(p_1+p_2+p_3)\Tr(\Taone_{U(N)}\Tatwo_{U(N)})
\Tr\, \T^{b_3}_{U(M)}\cr
&\qquad\qquad\qquad\Big[
e^{\iteta(p_1,p_2)}\triangle^{(1)-}_{\mu_1\mu_2}(p_1,p_2|\tilde{p}_3)+
e^{-\iteta(p_1,p_2)}\triangle^{(2)-}_{\mu_1\mu_2}(p_1,p_2|\tilde{p}_3) \Big],
\cr
&p_3^{\mu_3}\,
\langle j^{a_1}_{\;\;\;\mu_1}(p_1)\,j^{b_2}_{\;\;\;\mu_2}(p_2)\,
j^{a_3}_{\;\;\;\mu_3}(p_3) \rangle_{\rm con}^{\rm eps}=
(2\pi)^4 \delta(p_1+p_2+p_3)\Tr(\Taone_{U(N)}\Tathree_{U(N)})\Tr\, 
\T^{b_2}_{U(M)}\cr
&\qquad\qquad\qquad\Big[
e^{-\iteta(p_1,p_2)}\triangle^{(1)-}_{\mu_1\mu_2}(p_1,p_2|\tilde{p}_2)+
e^{\iteta(p_1,p_2)}\triangle^{(2)-}_{\mu_1\mu_2}(p_1,p_2|\tilde{p}_2)
\big],\cr
&p_3^{\mu_3}\,
\langle j^{b_1}_{\;\;\;\mu_1}(p_1)\,j^{a_2}_{\;\;\;\mu_2}(p_2)\,
j^{b_3}_{\;\;\;\mu_3}(p_3) \rangle_{\rm con}^{\rm eps}=-
(2\pi)^4 \delta(p_1+p_2+p_3)\Tr(\T^{b_1}_{U(M)}\T^{b_3}_{U(M)})
\Tr\, \Tatwo_{U(N)}\cr
&\qquad\qquad\qquad\Big[
e^{\iteta(p_1,p_2)}\triangle^{(1)+}_{\mu_1\mu_2}(p_1,p_2|\tilde{p}_2)+
e^{-\iteta(p_1,p_2)}\triangle^{(2)+}_{\mu_1\mu_2}(p_1,p_2|\tilde{p}_2)\big].
&\numeq\cr
}
\namelasteq{\nonplanaranomalous}
$$

From eq.~\anomalousplan, one deduces that the anomaly cancellation 
condition for the planar triangle  diagrams reads
$$
\Tr(\T^{a_1}_{U(N)}\T^{a_2}_{U(N)}\T^{a_3}_{U(N)})=0\qquad{\rm and}\qquad
\Tr(\T^{b_1}_{U(M)}\T^{b_2}_{U(M)}\T^{b_3}_{U(M)})=0.
$$
Both the anomalies which gives rise to these anomaly cancellation 
conditions are analogous to the anomaly in eq.~\triangleanomaly, i.e., the
anomaly for chiral fundamental fermions. If we now substitute 
eq.~\itisnonzero\ in eq~\nonplanaranomalous, we shall 
conclude that the left hand sides of last two identities in 
eq.~\bianomalyfreedom\ do not vanish, but read, respectively, thus 
$$
\eqalignno{
&p_3^{\mu_3}\,
\langle j^{a_1}_{\;\;\;\mu_1}(p_1)\,j^{b_2}_{\;\;\;\mu_2}(p_2)\,
j^{a_3}_{\;\;\;\mu_3}(p_3) \rangle_{\rm con}^{\rm eps}=
(2\pi)^4 \delta(p_1+p_2+p_3)\Tr(\Taone_{U(N)}\Tathree_{U(N)})\Tr\, 
\T^{b_2}_{U(M)}\cr
&\qquad\qquad\qquad\qquad\Big[
{1\over 2\pi^2}\sin{1\over 2}\theta(p_1,p_2)\;
\varepsilon_{\mu_1\mu_2\alpha\beta}\, {\tilde{p}_2^{\alpha}p_2^{\beta}
\over \tilde{p}_2^2}\cr
&\qquad\qquad\int_{0}^{1}\,dx\,\sqrt{\tilde{p}_2^2(-p_2^2-i0^+)x(1-x)}\;\;
{\rm K}_1\bigg(\sqrt{\tilde{p}_2^2(-p_2^2-i0^+)x(1-x)}\bigg)
\Big],\cr
&p_3^{\mu_3}\,
\langle j^{b_1}_{\;\;\;\mu_1}(p_1)\,j^{a_2}_{\;\;\;\mu_2}(p_2)\,
j^{b_3}_{\;\;\;\mu_3}(p_3) \rangle_{\rm con}^{\rm eps}=-
(2\pi)^4 \delta(p_1+p_2+p_3)\Tr(\T^{b_1}_{U(M)}\T^{b_3}_{U(M)})
\Tr\, \Tatwo_{U(N)}\cr
&\qquad\qquad\qquad\qquad\Big[
{1\over 2\pi^2}\sin{1\over 2}\theta(p_1,p_2)\;
\varepsilon_{\mu_1\mu_2\alpha\beta}\, {\tilde{p}_2^{\alpha}p_2^{\beta}
\over \tilde{p}_2^2}\cr
&\qquad\qquad\int_{0}^{1}\,dx\,\sqrt{\tilde{p}_2^2(-p_2^2-i0^+)x(1-x)}\;\;
{\rm K}_1\bigg(\sqrt{\tilde{p}_2^2(-p_2^2-i0^+)x(1-x)}\bigg)\Big].
&\numeq\cr
}
\namelasteq{\nonplanaranomaly}
$$
Recall that $\tilde{p}_2^{\mu}=\theta^{\mu\nu}p_{2\,\nu}$ and 
$\tilde{p}^2_2\equiv p_{2\,\mu}\,\theta^{\mu\rho}
\eta_{\rho\sigma}\theta^{\sigma\nu}\,p_{2\,\nu}$, so that
$\tilde{p}^2_2\geq 0$.

Finally, eqs.~\itvanishes\ imply  that
$$
p_3^{\mu_3}\,
\langle j^{a_1}_{\;\;\;\mu_1}(p_1)\,j^{a_2}_{\;\;\;\mu_2}(p_2)\,
j^{b_3}_{\;\;\;\mu_3}(p_3) \rangle_{\rm con}^{\rm eps}= 0.
$$
Similarly,
$$
p_3^{\mu_3}\,
\langle j^{b_1}_{\;\;\;\mu_1}(p_1)\,j^{b_2}_{\;\;\;\mu_2}(p_2)\,
j^{a_3}_{\;\;\;\mu_3}(p_3) \rangle_{\rm con}^{\rm eps}=0.
$$
To show that indeed the last four identities in eq.~\bianomalyfreedom\ hold, 
all that remains for us to do is to work out the following expressions
$$
\eqalign{
-e^{\iteta(p_1,p_2)}\,
\T^{a_1\,\io}_{\quad\ih}\,&\T^{a_3\, \ih}_{\quad\id}\,
\langle j^{(\A)\,\id}_{\;\;\;\mu_1\,\io}(-p_2)\,j^{b_2}_{\;\;\;\mu_2}(p_2)\,
\rangle_{\rm con}^{\rm eps}\cr
&+e^{-\iteta(p_1,p_2)}
\T^{a_3\, \jo}_{\quad\jh}\,\T^{a_1\,\jh}_{\quad\jd}\,
\langle j^{(\A)\,\jd}_{\;\;\;\mu_1\,\jo}(-p_2)\,j^{b_2}_{\;\;\;\mu_2}(p_2)\,
\rangle_{\rm con}^{\rm eps},\cr
-e^{\iteta(p_1,p_2)}\,
\T^{b_1\,\jo}_{\quad\jh}\,&\T^{b_3\, \jh}_{\quad\jd}\,
\langle j^{(\B)\,\jd}_{\;\;\;\mu_1\,\jo}(-p_2)\,j^{a_2}_{\;\;\;\mu_2}(p_2)\,
\rangle_{\rm con}^{\rm eps}\cr
&+e^{-\iteta(p_1,p_2)}
\T^{b_3\, \jo}_{\quad\jh}\,\T^{b_1\,\jh}_{\quad\jd}\,
\langle j^{(\B)\,\jd}_{\;\;\;\mu_1\,\jo}(-p_2)\,j^{a_2}_{\;\;\;\mu_2}(p_2)\,
\rangle_{\rm con}^{\rm eps}.\cr
}
$$
It is not difficult to see that the previous expressions are equal to
$$
\eqalignno{
&-i\,\Tr(\Taone_{U(N)}\Tathree_{U(N)})\Tr\, 
\T^{b_2}_{U(M)}\;
\sin{1\over 2}\theta(p_1,p_2)\;
\int {dq^4\over (2\pi)^2}\,e^{-i\theta(q,p_2)}\,
{{\rm tr}\{ (\qslash+\ptwoslash)\gamma_{\mu_2}\qslash\gamma_{\mu_1}\gamma_5\}
\over (q^2+i0^{+})((q+p_2)^2+i0^{+})},\cr
&-i\Tr(\T^{b_1}_{U(M)}\T^{b_3}_{U(M)})
\Tr\, \Tatwo_{U(N)}\;
\sin{1\over 2}\theta(p_1,p_2)\;
\int {dq^4\over (2\pi)^2}\,e^{i\theta(q,p_2)}\,
{{\rm tr}\{ (\qslash+\ptwoslash)\gamma_{\mu_2}\qslash\gamma_{\mu_1}\gamma_5\}
\over (q^2+i0^{+})((q+p_2)^2+i0^{+})},\cr
}
$$
respectively. Some algebra and the help of the appendix makes it possible
for us to conclude that the right hand sides of the last two identities 
in eq.~\bianomalyfreedom\ agree, respectively, with their left hand sides,  
the latter being given in eq.~\nonplanaranomaly.

In summary, we have shown that the last four identities of 
eq.~\bianomalyfreedom\ indeed hold in the quantum theory. 
These identities are the Ward identities
for the nonplanar contributions  to the three-point function of
$\Gamma[\A,\B]$: the Ward identities for the nonplanar triangle 
contributions. Hence, the nonplanar triangle contributions give rise to no 
gauge anomaly. On the other hand, the planar triangle contributions are
anomalous with  anomalies given in  eq.~\anomalousplan.

\section{3. The IR origin of  nonabelian chiral gauge anomalies}

In the previous section we have shown that, for the theories we are 
discussing, only planar triangle diagrams gives rise
to a gauge anomaly and we have given an UV interpretation of this anomaly.
Eq.~\triangleanomaly\ is the basic building-block for this type of 
anomaly: see eq.~\anomalousplan. To interpret the nonabelian chiral 
anomaly under scrutiny  as an IR phenomenon, we shall follow 
Coleman and Grossman~\cite{\ColemanGrossman} and compute 
$\Gamma_{\mu_1\mu_2\mu_3}^{a_1 a_2 a_3}(p_1,p_2)^{\eps}$ at the point
$$
p_1^2=p_2^2=p_3^2=-Q^2,\; p_1+p_2+p_3=0.
$$
$\Gamma_{\mu_1\mu_2\mu_3}^{a_1 a_2 a_3}(p_1,p_2)^{\eps}$ is the pseudotensor
part of the three-point function for a noncommutative gauge theory 
with a right-handed   fundamental fermion. The action
of this theory is given in eq.~\fundamentalaction. The corresponding IR
analysis for the planar triangle diagrams arising in the other theories 
studied in this paper (see eqs.
~\antifundamentalaction,~\adjointaction\ and~ \biaction) can be readily done
by adapting the results presented in the sequel.
  
Let us recall first that formally 
$\Gamma_{\mu_1\mu_2\mu_3}^{a_1 a_2 a_3}(p_1,p_2)^{\eps}$
is given by the sum of the pseudotensor contributions coming from the
triangle diagrams in fig. 2, which for the case at hand reads
$$
\eqalignno{\Gamma_{\mu_1\mu_2\mu_3}^{a_1 a_2 a_3}(p_1,p_2)^{\eps}=
&e^{-\iteta(p_1,p_2)}\,{\rm Tr}\,\Taone\Tatwo\Tathree
\,\triangle^{(1)}_{\mu_1\mu_2\mu_3}(p_1,p_2)+\cr
&e^{\iteta(p_1,p_2)}\,{\rm Tr}\,\Tatwo\Taone\Tathree
\,\triangle^{(2)}_{\mu_1\mu_2\mu_3}(p_1,p_2), 
&\numeq\cr
}
\namelasteq{\triangleonePirr}
$$
where
$$
\eqalign{&\triangle^{(1)}_{\mu_1\mu_2\mu_3}(p_1,p_2)\,=\,
\int{d^4 q\over(2\pi)^4}
{{\rm tr}^{\,\eps}\big\{ (\qslash+\poneslash)\,\gamma_{\mu_1}{\rm P}_{+}\,
\qslash\,\gamma_{\mu_2}{\rm P}_{+}\,(\qslash-\ptwoslash)\,\gamma_{\mu_3}{\rm P}_{+}\big\}
\over (q^2+i0^+) ((q+p_1)^2+i0^+) ((q-p_2)^2+i0^+)},\cr
&{\rm and}\cr
&\triangle^{(2)}_{\mu_1\mu_2\mu_3}(p_1,p_2)\,=\,
\int{d^4 q\over(2\pi)^4}
{{\rm tr}^{\,\eps}\big\{ (\qslash+\ptwoslash)\,\gamma_{\mu_2}{\rm P}_{+}\,
\qslash\,\gamma_{\mu_1}{\rm P}_{+}\,(\qslash-\poneslash)\,\gamma_{\mu_3}{\rm P}_{+}\big\}
\over (q^2+i0^+) ((q+p_2)^2+i0^+) ((q-p_1)^2+i0^+)}.
} 
$$
The symbol ${\rm tr}^{\,\eps}$ denotes the pseudotensor contributions, i.e.,
contributions involving an odd number of $\gamma_5$ matrices.

As they stand the Feynman amplitudes 
$\triangle^{(1)}_{\mu_1\mu_2\mu_3}(p_1,p_2)$ and
$\triangle^{(2)}_{\mu_1\mu_2\mu_3}(p_1,p_2)$ above are at first sight 
formal expressions since they are sum  of UV divergent 
by power-counting Feynman integrals. However, we shall see in a moment 
that one can associate to these Feynman amplitudes a unique 
tempered distribution provided cyclicity of the external indices and
momenta is imposed. Indeed,  renormalization theory~\cite{\RT}  
associates to every formal Feynman  amplitude  a tempered distribution 
which is uniquely defined up to a local polynomial of the appropriate 
dimension in the external momenta\footnote{*}{Here we assume that, since the
diagrams we are considering are one-loop and planar, 
standard renormalization theory can be applied to each diagram
without further ado.}. 
This polynomial can be further 
restricted by symmetries. Hence, the Feynman amplitude 
$\triangle^{(1)}_{\mu_1\mu_2\mu_3}(p_1,p_2)$ can 
be uniquely defined as a distribution modulo the following polynomial
$$
C_1\,\varepsilon_{\mu_1\mu_2\mu_3\alpha}\,p_1^{\alpha}\,+\,C_2\,
\varepsilon_{\mu_1\mu_2\mu_3\alpha}\,p_2^{\alpha},
\eqno\numeq\namelasteq\counterterm
$$       
where $C_1$ and $C_2$ are arbitrary constants. If we next impose symmetry
under cyclic permutations of the pairs 
$(\mu_1,p_1), (\mu_2,p_2), (\mu_3,p_3)$, with $p_1+p_2+p_3=0$, then  
$C_1$ and $C_2$ are fixed for once and all. Indeed, any further addition
ought to be of the type
$$
C_3\,\varepsilon_{\mu_1\mu_2\mu_3\alpha}\,(p_1\,+\, 
p_2\,+\,p_3)^{\alpha},
$$
which vanishes upon imposing four-momentum conservation. Actually, what
this discussion is telling us is that if we use, as intermediate computational 
procedure,  a regularization method that explicitly preserves 
the formal symmetry of $\triangle^{(1)}_{\mu_1\mu_2\mu_3}(p_1,p_2)$ 
under cyclic permutations of the pairs $(\mu_1,p_1)$, $(\mu_2,p_2)$, 
$(\mu_3,p_3)$, the limit in which the regulator is removed is well-defined. 
Besides, this limit is the same for all  regularizations 
(and, of course, renormalizations) of 
$\triangle^{(1)}_{\mu_1\mu_2\mu_3}(p_1,p_2)$ which preserve its formal cyclic
symmetry. Of course, any renormalization which breaks this cyclic 
symmetry can be brought to the unique symmetric form just 
mentioned by adding a finite counterterm of the form 
given in eq.~\counterterm . It is in this sense that we are entitled  
to say that the Feynman amplitude 
$\triangle^{(1)}_{\mu_1\mu_2\mu_3}(p_1,p_2)$ is an UV finite quantity, 
in spite of the fact that it is not UV finite by power-counting.
The same kind of arguments can be applied to 
$\triangle^{(2)}_{\mu_1\mu_2\mu_3}(p_1,p_2)$ to  conclude that it is also
an UV finite object, though it is not UV finite by power-counting. 

There is a very handy regularization procedure which explicitly preserves
the  symmetry of each triangle diagram in fig. 2  under cyclic permutations
of its external legs. This is the dimensional regularization algorithm set up
in the previous section. The dimensionally regularized counterparts of
$\triangle^{(1)}_{\mu_1\mu_2\mu_3}(p_1,p_2)$ and 
$\triangle^{(2)}_{\mu_1\mu_2\mu_3}(p_1,p_2)$ read:
$$
\eqalignno{&\triangle^{(1)}_{\mu_1\mu_2\mu_3}(p_1,p_2;d)\,=\,
\int{d^d q\over(2\pi)^d}
{{\rm tr}^{\,\eps}\big\{ (\qslash+\poneslash)\,\gambar_{\mu_1}{\rm P}_{+}\,
\qslash\,\gambar_{\mu_2}{\rm P}_{+}\,(\qslash-\ptwoslash)\,\gambar_{\mu_3}{\rm P}_{+}\big\}
\over (q^2+i0^+) ((q+p_1)^2+i0^+) ((q-p_2)^2+i0^+)},\cr
&{\rm and}\cr
&\triangle^{(2)}_{\mu_1\mu_2\mu_3}(p_1,p_2;d)\,=\,
\int{d^d q\over(2\pi)^d}
{{\rm tr}^{\,\eps}\big\{ (\qslash+\ptwoslash)\,\gambar_{\mu_2}{\rm P}_{+}\,
\qslash\,\gambar_{\mu_1}{\rm P}_{+}\,(\qslash-\poneslash)\,\gambar_{\mu_3}{\rm P}_{+}\big\}
\over (q^2+i0^+) ((q+p_2)^2+i0^+) ((q-p_1)^2+i0^+)}.
&\numeq\cr
} 
\namelasteq{\undimtrian}
$$
Taking into account that 
$$
{\rm P}_+\gamhat_{\mu}\gambar_{\nu}{\rm P}_+ =0,\;
{\rm P}_+\gambar_{\mu}\gambar_{\nu}=
\gambar_{\mu}\gambar_{\nu}{\rm P}_+ ,
$$
we conclude that eq.~\undimtrian\ can be turned into the following one
$$
\eqalignno{&\triangle^{(1)}_{\mu_1\mu_2\mu_3}(p_1,p_2;d)\,=\,
{1\over 2}\int{d^d q\over(2\pi)^d}
{{\rm tr}\big\{ (\bar{\qslash}+\bar{\poneslash})\,\gambar_{\mu_1}\,
\bar{\qslash}\,\gambar_{\mu_2}\,(\bar{\qslash}-\bar{\ptwoslash})\,\gambar_{\mu_3}\gamma_5\big\}
\over (q^2+i0^+) ((q+p_1)^2+i0^+) ((q-p_2)^2+i0^+)},\cr
&{\rm and}\cr
&\triangle^{(2)}_{\mu_1\mu_2\mu_3}(p_1,p_2;d)\,=\,
{1\over 2}\int{d^d q\over(2\pi)^d}
{{\rm tr}\big\{ (\bar{\qslash}+\bar{\ptwoslash})\,\gambar_{\mu_2}\,
\bar{\qslash}\,\gambar_{\mu_1}\,(\bar{\qslash}-\bar{\poneslash})\,\gambar_{\mu_3}\gamma_5\big\}
\over (q^2+i0^+) ((q+p_2)^2+i0^+) ((q-p_1)^2+i0^+)}.
&\numeq\cr
} 
\namelasteq{\almostthere}
$$
The computation of the previous integrals at $p_1^2=p_2^2=p_3^2=-Q^2$ is very
easy. The substitution in eq.~\almostthere\ of the integrals in the appendix
and some self-evident algebraic arrangements yield 
upon taking the limit $d\rightarrow 4$ the following result
$$
\eqalignno{\triangle^{(1)}_{\mu_1\mu_2\mu_3}(p_1,p_2)\,=&\,\
\triangle^{(2)}_{\mu_1\mu_2\mu_3}(p_1,p_2)\cr
={1\over 24\pi^2}\bigg({1\over Q^2}\bigg)
\big(\,\varepsilon_{\mu_1\mu_2\alpha\beta}\, 
p_1^{\alpha}p_2^{\beta}\,p_{3\,\mu_3}&+\varepsilon_{\mu_3\mu_1\alpha\beta}\, 
p_3^{\alpha}p_1^{\beta}\,p_{2\,\mu_2}+\varepsilon_{\mu_2\mu_3\alpha\beta}\, 
p_2^{\alpha}p_3^{\beta}\,p_{1\,\mu_1}\big).&\numeq\cr
}
\namelasteq{\IRpole}
$$
The Feynman amplitudes in the previous equation have poles at $Q^2=0$ 
and they are these IR singularities which we shall   hold responsible 
for the existence of the nonabelian chiral anomaly~\cite{\ColemanGrossman}.
If we now substitute eq.~\IRpole\ into eq.~\triangleonePirr\ we will obtain
the whole anomalous contribution to the three-point function at
$p_1^2=p_2^2=p_3^2=-Q^2$:
$$
\eqalignno{&\Gamma_{\mu_1\mu_2\mu_3}^{a_1 a_2 a_3}(p_1,p_2)^{\eps}\,=\,\cr
{1\over 24\pi^2}\,\bigg({1\over Q^2}\bigg)\,
\bigg({\rm Tr}\,\{\Taone,\Tatwo\}&\,\Tathree\cos \teta(p_1,p_2)
-i{\rm Tr}\,[\Taone,\Tatwo]\Tathree\sin\teta(p_1,p_2)\bigg)\cr
\big(\,\varepsilon_{\mu_1\mu_2\alpha\beta}\, 
p_1^{\alpha}p_2^{\beta}\,p_{3\,\mu_3}&+\varepsilon_{\mu_3\mu_1\alpha\beta}\, 
p_3^{\alpha}p_1^{\beta}\,p_{2\,\mu_2}+\varepsilon_{\mu_2\mu_3\alpha\beta}\, 
p_2^{\alpha}p_3^{\beta}\,p_{1\,\mu_1}\big).
&\numeq\cr
}
\namelasteq{\anomalousthreept}
$$
Notice that by contracting with $p_3^{\mu_3}$ both sides of the 
previous equation, one obtains once again the anomaly equation 
(eq.~\triangleanomaly\ ). Also notice that unlike in the commutative case
the r.h.s of eq.~\anomalousthreept\ vanishes if and only if
${\rm Tr}\,\{\Taone,\Tatwo\}\,\Tathree=0$ and 
${\rm Tr}\,[\Taone,\Tatwo]\Tathree=0$, i.e., 
${\rm Tr}\; \Taone\Tatwo \Tathree=0$. Indeed, the nonpolynomial -in the
Moyal product- IR contributions, 
$$
{\cos \teta(p_1,p_2)\over Q^2}\quad{\rm and}\quad
{\sin \teta(p_1,p_2)\over Q^2},
$$
in this equation, makes it impossible for us to redefine 
$\Gamma_{\mu_1\mu_2\mu_3}^{a_1 a_2 a_3}(p_1,p_2)^{\eps}$ so that the anomaly
cancellation condition read merely ${\rm Tr}\,\{\Taone,\Tatwo\}\,\Tathree=0$.

\section{4. Summary and Conclusions}

In this paper we have shown that the one-loop noncommutative nonabelian 
gauge anomalies for $U(N)$ groups can be interpreted either as an UV effect 
or as an IR phenomenon. We have considered three basic types of 
noncommutative chiral gauge theories, namely, gauge theories with a
fundamental, gauge theories with an adjoint and gauge theories with a 
bi-fundamental right-handed fermion. We have computed the  anomaly in
one-loop planar triangle diagrams and shown that the nonplanar    
contributions yield no gauge anomaly since they preserve the corresponding
Ward identities. It turned out that chiral gauge theories with fundamental, 
anti-fundamental and bi-fundamental  matter are, in general,  
anomalous and that chiral theories with only adjoint fermions are 
-due to a special cancellation mechanism- always anomaly free. Last but not
least, we have clarified the origin of the noncommutative anomaly cancellation
condition $\Tr\,\Taone\Tatwo\Tathree=0$.   

It will be interesting to carry out the analysis presented
here for the theories introduced in ref.~\cite{\Bonoraetal} and
for the axial anomaly~\cite{\Arsa}. Anomalies in the presence of 
noncommutative gravity~\cite{\Gravity} are also worth studying. We shall
report on these topics elsewhere.

\bigskip

\section{ Appendix}

The following result is needed to obtain eq.~\itisnonzero:
$$
\eqalign{
\int{d^4 q\over(2\pi)^4}\,
e^{\pm i\theta(q,p)}\;&
{q^{\mu}\over (q^2+i0^+)((q+p)^2+i0^+)}=\cr
&-{i\,p^{\mu}\over 8\pi^2}\,\int_{0}^{1}\,dx\,x\;
{\rm K}_0\bigg(\sqrt{\tilde{p}^2(-p^2-i0^+)x(1-x)}\bigg)\cr
\pm\,{1\over 8\pi^2}&{\tilde{p}^{\mu} \over \tilde{p}^2}
\,\int_{0}^{1}\,dx\,\sqrt{\tilde{p}^2(-p^2-i0^+)x(1-x)}\;\;
{\rm K}_1\bigg(\sqrt{\tilde{p}^2(-p^2-i0^+)x(1-x)}\bigg),\cr
}
$$
where $\tilde{p}^{\mu}=\theta^{\mu\nu}p_{\nu}$, but 
$\tilde{p}^2\equiv p_{\mu}\,\theta^{\mu\rho}
\eta_{\rho\sigma}\theta^{\sigma\nu}\,p_{\nu}$, so that
$\tilde{p}^2_i\geq 0$.

Next, we display the integrals needed to obtain eq.~\IRpole . These integrals
are worked out at the point $p_1^2=p_2^2=-2\,p_1\cdot p_2=-\,Q^2$. They read
$$
\eqalign{
\int {d^d q\over(2\pi)^d}\;{1\over q^2\,(q+p_1)^2\,(q-p_2)^2}\,=\, &{\Phi\over
Q^2}\,+\, O(d-4),\cr
\int {d^d q\over(2\pi)^d}\;{\bar{q}_{\mu}\over q^2\,(q+p_1)^2\,(q-p_2)^2}\,
=\, &\bigg({\Phi\over 3}\bigg)
\bigg({1\over Q^2}\bigg)(\bar{p}_2-\bar{p}_1)_{\mu}\,+\, O(d-4), \cr
\int {d^d q\over(2\pi)^d}\;
{\bar{q}_{\mu}\bar{q}_{\nu}\over q^2\,(q+p_1)^2\,(q-p_2)^2}\,
=\, &
\bigg({{\rm I_1}\over 4}+{\Phi\over 6} + 
{{\rm I_2}\over 4}\bigg)\, \gbar_{\mu\nu}
+{{\rm I_2}\over 6}\bigg({1\over Q^2}\bigg)
\big(\bar{p}_{1\,\mu}\bar{p}_{2\,\nu}+\bar{p}_{2\,\mu}\bar{p}_{1\,\nu}\big)+\cr
&\bigg({\Phi\over 3} + {{\rm I_2}\over 3}\bigg)\bigg({1\over Q^2}\bigg)
\big(\bar{p}_{1\,\mu}\bar{p}_{1\,\nu}+\bar{p}_{2\,\mu}\bar{p}_{2\,\nu}\big)
\,+\, O(d-4), \cr
\int {d^d q\over(2\pi)^d}\;{\bar{q}^2\,\bar{q}_{\mu}\over 
q^2\,(q+p_1)^2\,(q-p_2)^2}\,
=\, &\bigg({{\rm I_1}\over 2}+{{\rm I_2}\over 6}\bigg)(\bar{p}_2-\bar{p}_1)_{\mu}\,+\, O(d-4), \cr
}
$$
where 
$$
\eqalign{
&{\rm I_1}\,=\,{i\over 16\pi^2}\,\bigg(-{1\over\epsilon}-\gamma-
\ln\,{Q^2\over 4\pi\kappa^2}+2\bigg),\cr
&{\rm I_2}\,=\,{i\over 16\pi^2},\cr
&\Phi\,=\,{i\over 16\pi^2}\int_{0}^1\,dx\;{\ln x(1-x)\over 1-x+x^2},\cr
}
$$
and $d\,=\,4+2\epsilon$.

Note that all the ugly features of the integrals above nicely cancel against
one another when substituted in eq.~\almostthere\ to yield the beautiful 
result of eq.~\IRpole.

\bigskip
\section{Acknowledgments}
This paper grew out of a suggestion made to me by Prof. L. Alvarez-Gaum\'e  
that chiral anomalies on noncommutative Minkowski space-time should  have 
an IR interpretation. I am indebted to  him for correspondence on this point. 
This work has been partially supported by CICyT under 
grant PB98-0842.

\bigskip\bigskip

\section{References}

\frenchspacing

\refno\elementos.
J.M. Gracia-Bond\'{\i}a, J.C. V\'arilly and H. Figeroa, {\it Elements of
Noncommutative Geometry}, Birkh\"auser, Boston, 2000.

\refno\AGM.
O. Aharony, J. Gomis and T. Mehen,
{\it On Theories with Light-Like Noncommutativity}, {\tt hep-th/0006236}.

\refno\Terashima.
S. Terashima, Phys. Lett. B482 (2000) 276.

\refno\anomalPP.
J.M. Gracia-Bond\'{\i}a and C.P. Mart\'{\i}n, Phys. Lett. B479 (2000) 312.

\refno\Bonora.
L. Bonora, M. Schnabl and A. Tomasiello, Phys. Lett. B485 (2000) 311.

\refno\Adler.
S. Adler, Phys. Rev. 177 (1969) 2426; J. Bell and R. Jackiw,
Nuovo Cimento A 60 (1969) 47; W. Bardeen, Phys. Rev. 184 (1969)
1848; S.L. Adler, {\it Perturbation Theory Anomalies}, 
Lectures on Elementary Particles and Quantum Field Theory, 
1970 Brandeis University Summer Institute in Theoretical Physics, 
The MIT Press.

\refno\ColemanGrossman.
A.D. Dolgov and V.I. Zakharov, Nucl. Phys.  B27 (1971) 525;
Y. Frishman, A. Schwimmer, T. Banks and S. Yankielowicz, Nucl. Phys. B177 (
1981) 157;
S. Coleman and B. Grossman, Nucl. Phys. B203 (1982) 205.

\refno\MRS.
S. Minwalla, M. Van Raamsdonk and N. Seiberg, J. High Energy Phys. 0002 (2000)
020.

\refno\Intanom.
L. Alvarez-Gaum\'e, {\it Introduction to Anomalies}, Erice School Math. Phys.
1985:0093.

\refno\BreitMaison.
P. Breitenlohner and D. Maison, Comm. Math. Phys. 52 (1977) 11, {\it ibid.}
52 (1977) 39, 52 (1977) 55.

\refno\Pernici.
M. Pernici, M. Raciti and F. Riva, Nucl. Phys. B577 (2000) 293;
M. Pernici, {\it Semi-Naive dimensional renormalization}, {\tt hep-th/9912278};
M. Pernici and M. Raciti, {\it Axial current in QED and 
semi-naive dimensional renormalization}, {\tt hep-th/0003062}.

\refno\gammafive.
C.P. Mart\'{\i}n and D. S\'anchez-Ruiz, Nucl. Phys. B572 (2000) 387.

\refno\AGO.
A. Gonz\'alez-Arroyo and 
M. Okawa, Phys. Lett. B120 (1983) 174, Phys. Rev. D27 (1983)
2397; T. Filk, Phys. Lett. B376 (1996) 53.

\refno\RT.
K. Hepp, {\it Renormalization Theory}, Statistical Mechanics and Quantum Field
Theory, Les Houches 1970, Gordon and Breach Science Publishers;
H. Epstein and V. Glaser, Ann. Inst. Henri Poincar\'e, XIX 3 (1973) 211;
P. Blanchard and R. Seneor, Ann. Inst. Henri Poincar\'e, XXIII 2 (1975) 147;
R. Brunetti and K. Fredenhagen, Comm. Math. Phys. 208 (2000) 623.

\refno\Bonoraetal.
L. Bonora, M. Schnabl, M.M. Sheikh-Jabbari and A. Tomasiello, {\it
Noncommutative $SO(n)$ and $Sp(n)$ Gauge Theories}, {\tt hep-th/0006091}.

\refno\Arsa.
F. Ardalan and N. Sadooghi, {\it Axial anomaly in noncommutative QED on
${\rm I\!R}^4$}, {\tt hep-th/0002143}.

\refno\Gravity.
A.H. Chamseddine, {\it Complexified Gravity in Noncommutative Spaces},
\hfil\break
{\tt hep-th/0005222}; J.W. Moffat, {\it Noncommutative Quantum Gravity},
{\tt hep-th/0007181}.

\end